\documentclass[a4paper,11pt]{article}
\pdfoutput=1 
\usepackage{jheppub} 
\usepackage{color}
\usepackage{graphicx, multirow,soul,amsmath,amsfonts,amssymb,mathrsfs,amsfonts}
\usepackage{hyperref}
\usepackage[all]{hypcap}
\usepackage{url}
\usepackage{bbm}
\usepackage{cancel}
\usepackage{slashed}
\usepackage[dvipsnames]{xcolor}
\usepackage{multicol, blindtext}
\usepackage{lipsum}
\usepackage{mattens}
\usepackage{mathtools}
\usepackage{footnote}
\usepackage{subcaption}
\usepackage[utf8]{inputenc}

\usepackage{braket}

\newcommand{\MeV}{\ensuremath{\rm~MeV}\xspace}
\newcommand{\GeV}{\ensuremath{\rm~GeV}\xspace}
\newcommand{\TeV}{\ensuremath{\rm~TeV}\xspace}

\newcommand{\be}{\begin{equation}}
\newcommand{\ee}{\end{equation}}

\newcommand{\beq}{\begin{equation}}
\newcommand{\eeq}{\end{equation}}
\newcommand{\hc}{\mathrm{h.c.}}

\newcommand{\bea}{\begin{eqnarray}}
\newcommand{\eea}{\end{eqnarray}}

\newcommand\snowmass{\begin{center}\rule[-0.2in]{\hsize}{0.01in}\\\rule{\hsize}{0.01in}\\
Submitted to the  Proceedings of the US Community Study\\ 
on the Future of Particle Physics (Snowmass 2021)
\vskip -0.09in
\rule{\hsize}{0.01in}\\\rule[+0.2in]{\hsize}{0.01in} \end{center}}

\begin{document} 
\title{New Ideas in Baryogenesis: 
A Snowmass White Paper}

\author[1]{Editors:  Gilly Elor,}
\author[2]{Julia Harz,}
\author[3]{Seyda Ipek,}
\author[4]{Bibhushan Shakya.\vspace{.3cm}} 
 \author[]{} \author[]{}\author[]{}\author[]{}\author[]{}\author[]{}\author[]{}\author[]{}\author[]{}\author[]{}\author[]{}\author[]{}\author[]{} \author[]{}\author[]{}\author[]{}\author[]{} \author[]{}\author[]{}\author[]{}\author[]{}\author[]{}\author[]{}\author[]{}\author[]{}\author[]{}\author[]{}\author[]{}\author[]{}\author[]{}
\author[5]{Authors: Nikita Blinov,}
\author[6]{Raymond T. Co,}
\author[7]{Yanou Cui,}
\author[8]{Arnab Dasgupta,}
\author[9]{Hooman Davoudiasl,}
\author[1]{Fatemeh Elahi,}
\author[1]{Gilly Elor,}
\author[2]{ K\aa re Fridell,}
\author[8]{Akshay Ghalsasi,}
\author[10]{Keisuke Harigaya,}
\author[2]{Julia Harz,}
\author[2]{Chandan Hati,}
\author[11]{Peisi Huang,}
\author[3]{Seyda Ipek,}
\author[10]{Azadeh Maleknejad,} 
\author[12]{Robert McGehee,}
\author[13]{David E. Morrissey,}
\author[10]{Kai Schmitz,}
\author[4]{Bibhushan Shakya,}
\author[13]{Michael Shamma,}
\author[14]{Brian Shuve,}
\author[15]{David Tucker-Smith,}
\author[4]{Jorinde van de Vis,} 
\author[16]{Graham White.}

\affiliation[1]{PRISMA$^+$ Cluster of Excellence $\&$ Mainz Institute for Theoretical Physics\\
Johannes Gutenberg University, 55099 Mainz, Germany}
\affiliation[2]{Physik Department T70, Technische Universität München,
James-Franck-Straße 1, 85748 Garching, Germany}
\affiliation[3]{Carleton University
1125 Colonel By Drive, Ottawa, Ontario K1S 5B6, Canada}
\affiliation[4]{{Deutsches Elektronen-Synchrotron DESY, Notkestrasse 85, 22607 Hamburg, Germany}}
\affiliation[5]{Department of Physics and Astronomy, University of Victoria, Victoria, BC V8P 5C2, Canada}
\affiliation[6]{William I. Fine Theoretical Physics Institute, School of Physics and Astronomy, University of Minnesota, Minneapolis, MN 55455, USA}
\affiliation[7]{Department of Physics and Astronomy, University of California, Riverside, CA 92521, USA}
\affiliation[8]{High Energy Theory Group, Physics Department, Brookhaven National Laboratory,
Upton, New York 11973, USA}
\affiliation[9]{
Pittsburgh Particle Physics, Astrophysics, and Cosmology Center, Department of Physics and Astronomy,
University of Pittsburgh, Pittsburgh, USA
}
\affiliation[10]{Theoretical Physics Department, CERN, 1211 Geneva 23, Switzerland}
\affiliation[11]{Department of Physics and Astronomy, University of Nebraska, Lincoln, NE 68588, USA}
\affiliation[12]{Leinweber Center for Theoretical Physics, Department of Physics,University of Michigan, Ann Arbor, MI 48109, USA}
\affiliation[13]{TRIUMF, Theory Department, 4004 Wesbrook Mall, Vancouver, BC V6T 2A3, Canada}
\affiliation[14]{Department of Physics, Harvey Mudd College, Claremont, CA 91711, USA}
\affiliation[15]{
Department of Physics, Williams College, Williamstown, MA 01267, USA}
\affiliation[16]{Kavli IPMU (WPI), UTIAS, The University of Tokyo, Kashiwa, Chiba 277-8583, Japan}

\snowmass{}

\abstract{
The Standard Model of Particle Physics cannot explain the observed baryon asymmetry of the Universe. This observation is a clear sign of new physics beyond the Standard Model. There have been many recent theoretical developments to address this question. Critically, many new physics models that generate the baryon asymmetry have a wide range of repercussions for many areas of theoretical and experimental particle physics.
This white paper provides an overview of such recent theoretical developments with an emphasis on experimental testability. 
\vskip 0.5in
} 

\makeatletter
\def\@fpheader{}
\makeatother
\toccontinuoustrue
\maketitle

\newpage
\section{Executive Summary}
\label{sec:intro}
There is more matter than antimatter in the Universe. This asymmetry, quantified as the ratio of baryon density to photon density, is measured at the time of Big Bang Nucleosynthesis (BBN) and the Cosmic Microwave Background (CMB) to be
$
 (n_b-n_{\bar{b}})/n_\gamma = n_b/n_\gamma = \left(6.10\pm 0.4\right)\times 10^{-10}\,
$~\cite{planck}.
Inflation dictates that such an asymmetry must be dynamically generated after reheating, necessitating a mechanism of \emph{baryogenesis}. 

\vspace{0.4cm}

\begin{figure}[h]
    \centering
    \includegraphics[width=.8\textwidth]{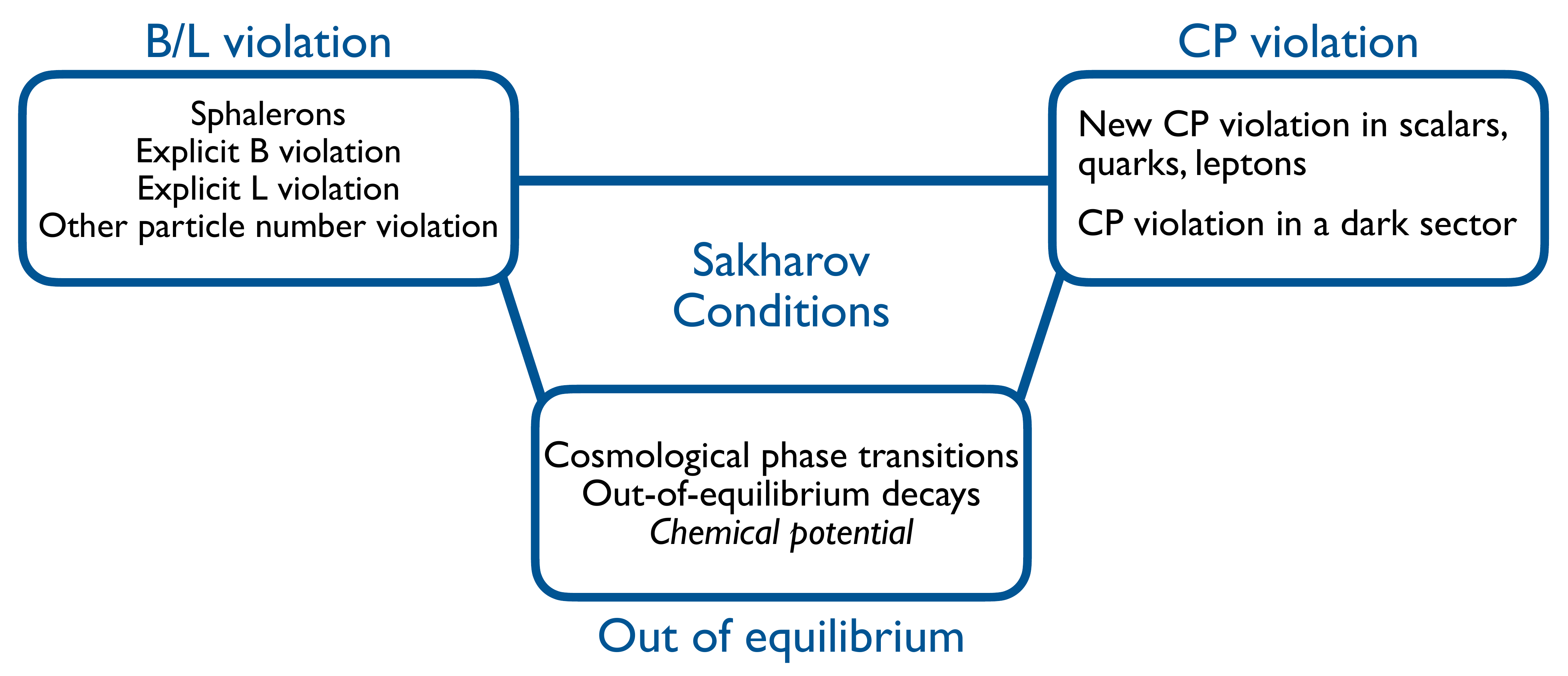}
    \caption{Some of the traditional BSM ingredients evoked to satisfy the Sakharov conditions \cite{sakharov} and explain the BAU. }
    \label{fig:BSMing}
\end{figure}

\vspace{0.4cm}

In order to produce a matter--antimatter asymmetry, a model of particle physics must satisfy the so-called Sakharov conditions \cite{sakharov}. These are: (i) Baryon number ($B$) violation, (ii) $C$ and $CP$ violation, and (iii) departure from thermodynamic equilibrium. 
In the Standard Model  (SM), (i) Baryon number is anomalously violated in the weak interactions of the SM. Although the rate of $B$-violating  \emph{sphaleron} processes is exponentially suppressed at zero temperature, sphalerons are very efficient at temperatures at which electroweak symmetry is restored, $T\gtrsim 130$~GeV \cite{Arnold:1987mh,Arnold:1996dy}. (ii) There is $CP$ violation in the CKM matrix, and possibly in the PMNS matrix~\cite{Abe:2019vii}. It has been argued that the CKM phase is not sufficient (in fact orders of magnitude too small) for producing the baryon asymmetry of the Universe (BAU). Within the SM there is no process to employ the $CP$ violation in the PMNS matrix to produce the BAU. (iii) There are many ways a process could occur out of thermal equilibrium, such as particle decay at temperatures below its mass, or a first-order phase transition. There is no process in the SM that goes out of thermal equilibrium in the early Universe. These shortcomings of the SM are a clear sign of BSM physics. By the nature of the problem, these observations and the related new physics have strong implications for early Universe cosmology. Beyond-the-Standard Model (BSM) models that seek to explain the BAU invoke certain ingredients to satisfy the Sakharov conditions (see Figure \ref{fig:BSMing} for a short summary).

\begin{figure}[h]
    \centering
    \includegraphics[width=.8\textwidth]{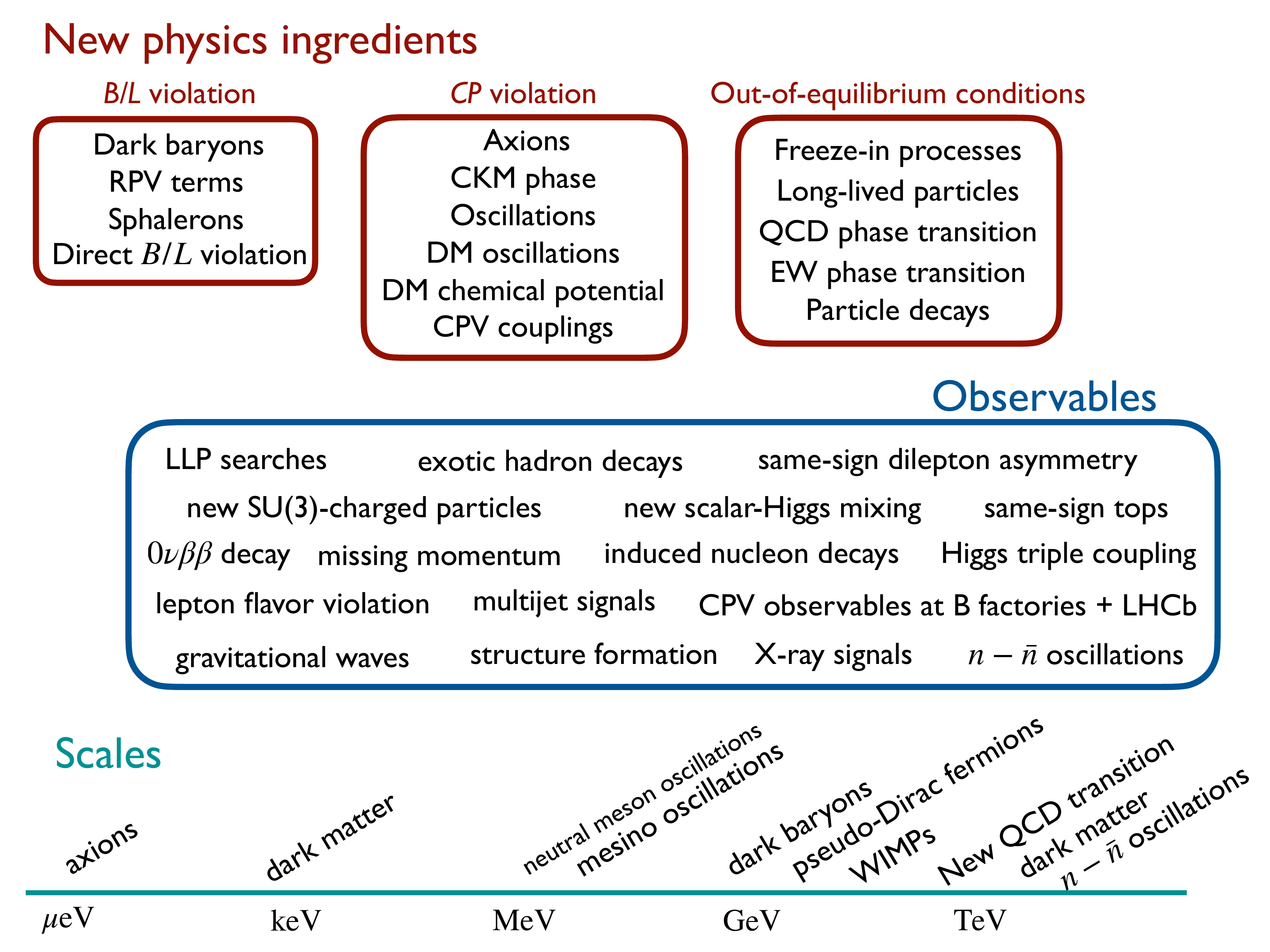}
    \caption{An incomplete but representative sample of new physics ingredients that are introduced to satisfy the Sakharov conditions, and various associated experimental observables that can arise in new proposed mechanisms of baryogenesis. Unlike traditional high scale mechanisms of baryogenesis, novel proposals lead to mechanisms that can produce the baryon asymmetry over a variety of scales.}
    \label{fig:newmodels}
\end{figure}

 Although the question of generating the BAU has been around for several decades, particle theorists are still coming up with novel ways to address this mystery, inspired by new theoretical ideas and observational opportunities. The goal of this white paper is to provide an overview of such proposals. Section~\ref{sec:NewModels} contains a collection of baryogenesis mechanisms that were proposed in the last decade, where contributing authors discuss the main ideas and experimental signals of each mechanism. This is not intended to be an exhaustive list, but simply a representation of the novel developments in the field. In Table \ref{tab:newmodels} we provide a brief summary of these models and their key points. A salient aspect of these contributions is experimental testability: while the vast majority of traditional baryogenesis models have involved high-scale physics and hence are difficult to probe experimentally, many new BSM models produce the BAU at low scales and involve low-scale new physics, and are therefore experimentally observable. Many of these new and exciting models are expected to produce signals at multiple experiments, allowing for a multi-prong search for new physics. In Figure \ref{fig:newmodels}, we illustrate the new physics ingredients, scales, and observables in these models. Section~\ref{sec:NewTests} summarizes recent proposals for searches that could shed light on new and some of the more traditional mechanisms of baryogenesis.

\begin{table}[t]
    \centering
    \begin{tabular}{l}
    \includegraphics[width=\textwidth]{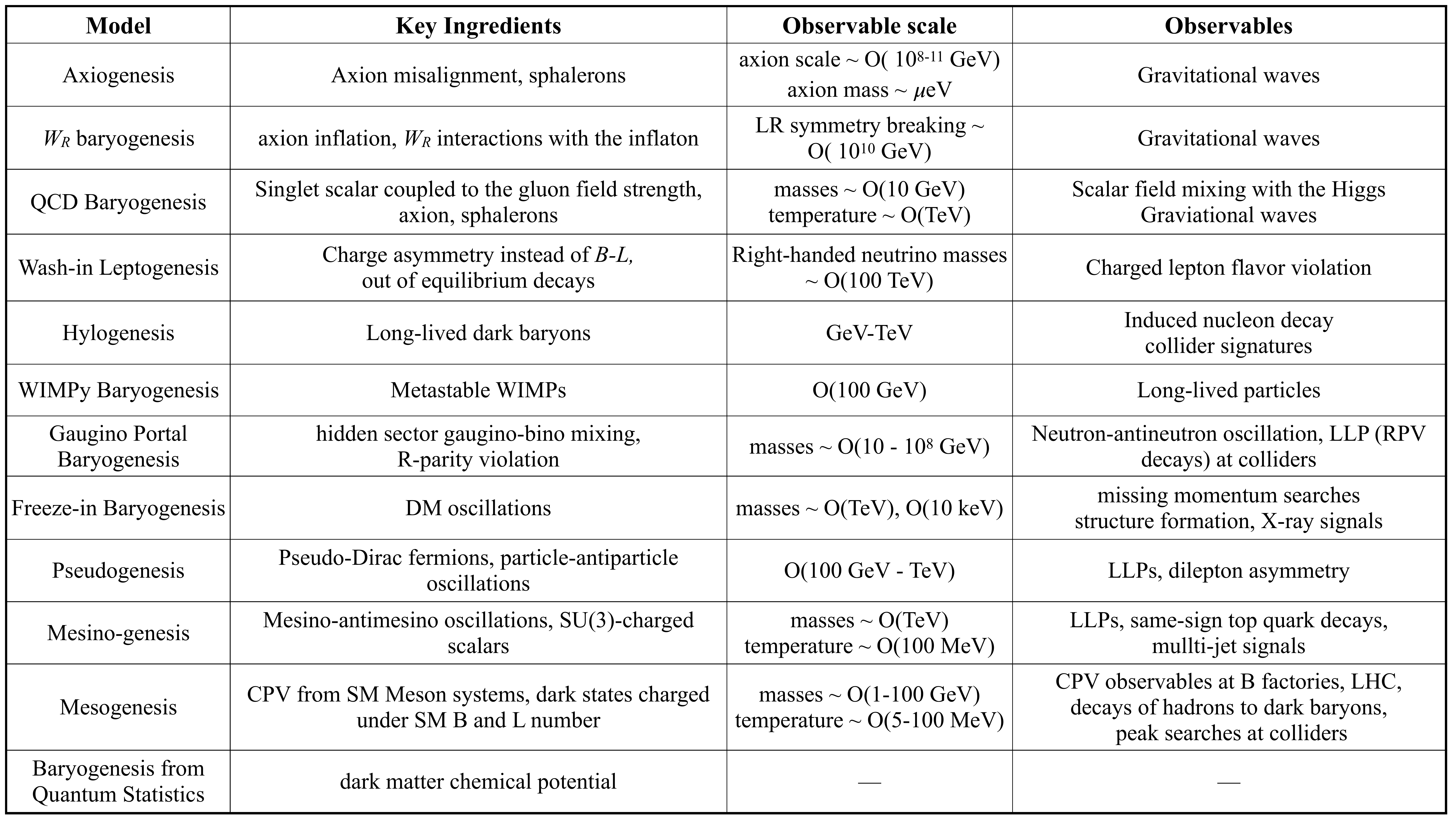}    \end{tabular}
    \caption{A brief summary of the new physics models surveyed in this white paper. The \emph{key ingredients} give an idea of the aspects of the model that lead to the BAU. Where available, we identified the new physics mass scales and the temperature at which the BAU could be generated. For details, see Section \ref{sec:NewModels}.}
    \label{tab:newmodels}
\end{table}

\newpage

\section{New Ideas in Baryogenesis Models}
\label{sec:NewModels}
 The question of generating the BAU has been around for decades and is one of the strongest drivers of new physics ideas. In order to showcase the exciting developments in this area, we reached out to several researchers who worked on such new physics models in the recent years. Below is a compilation of such models, many of them with exciting experimental implications. This is of course an incomplete list, but it points to the several new directions the field has taken.


\subsection{Axiogenesis}
\subsection*{Contributors: Raymond Co and Keisuke Harigaya}

Axions are highly motivated as new physics beyond the SM. The QCD axion~\cite{Weinberg:1977ma,Wilczek:1977pj} arises from the Peccei-Quinn ($PQ$) symmetry~\cite{Peccei:1977hh, Peccei:1977ur} that solves the strong $CP$ problem. The axion-like particles (ALPs) appear in other spontaneously broken global $U(1)$ symmetries that solve various problems in the SM such as the flavor symmetry~\cite{Froggatt:1978nt} or the lepton symmetry~\cite{Chikashige:1980ui}. Such axions in the light mass regime are cosmologically stable and therefore serve as excellent dark matter candidates~\cite{Preskill:1982cy, Dine:1982ah,Abbott:1982af}.

The axiogenesis mechanism~\cite{Co:2019wyp} proposes that the observed baryon asymmetry of the universe can be generated by the novel axion field dynamics in the early universe in a way that the dark matter abundance is explained by the same dynamics.
In field theoretical realizations of axions, the axion field may obtain a nonzero field velocity from the explicit $PQ$ symmetry breaking higher-dimensional potential $\sim P^n$, where the complex field $P = S e^{i a / S} / \sqrt{2}$ contains both the radial mode $S$ and the axion $a$. Such explicit $PQ$ breaking is effective when the radial mode $S$ takes on a large field value in the early universe and can cause $P$ to rotate in field space when the Hubble friction is overcome by the gradient of the $PQ$-conserving potential. This way of the initiation of rotations of complex scalar fields was proposed in the Affleck-Dine mechanism~\cite{Affleck:1984fy}. The rotation corresponds to a conserved $PQ$ charge $n_{\rm PQ} = S^2 \dot\theta $ with $\dot\theta \equiv \dot a / S$.

This $PQ$ charge can be partially transferred to the SM chiral asymmetries via the axion couplings with SM particles. Baryon number-violating processes, such as the electroweak sphaleron process, can generate the baryon asymmetry from the chiral asymmetries. 

The kinetic energy of the remaining axion rotation can provide the abundance for an axion with mass $m_a$,
\begin{align}
    \frac{\rho_a}{s} \simeq 2 m_a \frac{n_{\rm PQ}}{s},
\end{align}
which can be larger than that from the potential energy as in the conventional misalignment mechanism~\cite{Preskill:1982cy, Dine:1982ah,Abbott:1982af}. This dark matter production mechanism is called the kinetic misalignment mechanism~\cite{Co:2019jts,Co:2020dya}.

The QCD axion, by definition, has a QCD anomaly. Therefore, the $PQ$ charge is transferred to the quark chiral asymmetries via the strong sphaleron processes.
Baryon number violation is then provided by electroweak sphaleron processes, by which the final baryon asymmetry is generated from the quark chiral asymmetries and freezes out at the electroweak phase transition
\begin{align}
    Y_B = \frac{n_B}{s} = \left. \frac{45 c_B}{2 g_* \pi^2} \frac{\dot{\theta}}{T} \right|_{T = T_{\rm EW}},
\end{align}
where $c_B$ is a constant that is typically $\mathcal{O}(0.1)$, but can be larger if the coupling of the axion with the SM fermions or the weak gauge boson is much larger than the coupling with the gluon.
In this minimal scenario presented in Ref.~\cite{Co:2019wyp}, the observed abundances of dark matter and the baryon asymmetry predict the temperature at which the electroweak sphaleron processes fall out of thermal equilibrium, $T_{\rm EW} \simeq 1~{\rm TeV} (f_a / 10^8 \GeV) (0.1/c_B)^{1/2}$, with $f_a$ the axion decay constant, whereas a lower $T_{\rm EW}$ leads to overproduction of axion dark matter. The SM prediction $T_{\rm EW} = 130$ GeV  is incompatible with the astrophysical lower bound $f_a > 10^8$ GeV unless $c_B > 10$. Higher electroweak phase transition temperature is possible by couplings of the SM Higgs with new particles~\cite{Ishikawa:2014tfa,Baldes:2018nel,Glioti:2018roy,Co:2019wyp}, and we may correlate the QCD axion decay constant with the mass scale of the new particles that can be measured by collider experiments. 

This picture is modified with the presence of $B-L$ violation at high temperatures since the electroweak sphaleron processes only violate $B+L$ and preserve $B-L$. Whenever new physics is involved in providing the extra $B-L$ violation, non-trivial constraints from dark matter and the baryon asymmetry will lead to predictions of new physics parameters. Examples of such $B-L$ violation and the associated predictions are as follows. Lepton number violation can originate from the dimension-5 Majorana neutrino mass terms as considered in Refs.~\cite{Domcke:2020kcp, Co:2020jtv, Kawamura:2021xpu}, which predicts scalar masses compatible with the mini-split supersymmetry~\cite{Co:2020jtv, Kawamura:2021xpu}. Dirac neutrino masses in composite theories can also produce effective $B-L$ asymmetry as considered in Ref.~\cite{Chakraborty:2021fkp}, where an observable amount of dark radiation is predicted. Baryon and/or lepton number violation can arise from $R$-parity violation~\cite{Co:2021qgl}, predicting the lifetime of the lightest supersymmetric particle that can be measured at colliders in some of the parameter space. Nucleon decays are also predicted under plausible assumptions.
An extra non-Abelian gauge symmetry $SU(2)_R$ can also give rise to baryon number violation~\cite{Harigaya:2021txz}, and this scenario predicts a new gauge boson that can be probed at the high luminosity LHC. All of these axiogenesis scenarios avoid overproduction of axion dark matter even with $T_{\rm EW} = 130 \GeV$ and prefer an axion decay constant $f_a = 10^{8\mathchar`-11} \GeV$, smaller than that of conventional predictions. 

The idea of axiogenesis can be extended to ALPs, in which the $PQ$ charge discussed above is replaced by the $U(1)$ charge of the ALPs. In the so-called ALP cogenesis~\cite{Co:2020xlh}, dark matter and the baryon asymmetry can be simultaneously explained by the ALP rotation without the need of a higher $T_{\rm EW}$ nor additional new physics. Due to the absence of the QCD anomaly, the $U(1)$ charge associated with the ALP rotations will not be transferred via the QCD sphaleron, but the transfer to particle-antiparticle asymmetry can proceed because of a possible electroweak anomaly and/or axion-fermion couplings. Such particle asymmetries will be reprocessed to the baryon asymmetry via electroweak sphaleron processes. Finally, the dark matter abundance is correctly reproduced by an ALP mass lighter than that of the QCD axion for a given $f_a$, and ALP cogenesis predicts a relation between $f_a$ and $m_a$,
\begin{align}
    f_a=2 \times 10^{9} {\GeV} \left(\frac{f_a}{S\left(T_{\rm EW}\right)}\right)\left(\frac{c_{B}}{0.1}\right)^{\frac{1}{2}}\left(\frac{\mu\mathrm{eV}}{m_{a}}\right)^{\frac{1}{2}}\left(\frac{T_{\rm EW}}{130 {\GeV}}\right) ,
\end{align}
which is more experimentally accessible than the QCD axion and than the prediction of the misalignment mechanism. Here the parameter $S\left(T_{\rm EW}\right)$, which is the effective decay constant at $T_{\rm EW}$, allows the possibility where the radial mode $S$ has not settled down to the minimum at $f_a$.

In all these axiogenesis scenarios involving the QCD axion or ALPs, the axion rotation can bring about a period of kination---a cosmological era where the total energy density of the universe is dominated by axion field's kinetic energy~\cite{Co:2019wyp}. Such a kination era can lead to enhanced signals of gravitational waves from inflation and/or cosmic strings with a unique triangular spectrum~\cite{Co:2021lkc, Gouttenoire:2021wzu, Gouttenoire:2021jhk}, by which the parameter space compatible with axiogenesis can be probed. A separate gravitational wave signal may also arise from the rotation itself in the presence of a coupling with a dark photon~\cite{Co:2021rhi,Madge:2021abk}. In addition, the rotation can reheat the universe via its thermal dissipation, in which case the rotation also serves as a curvaton, sourcing the observed cosmic perturbations with a likely detectable amount of non-Gaussianity~\cite{Co:2022qpr}.

\subsection{$\rm{W_R}$-Axion Baryogenesis and Darkgenesis}
\subsection*{Contributor: Azadeh Maleknejad}

\begin{figure}[h]
    \centering
    \includegraphics[width=0.9\textwidth]{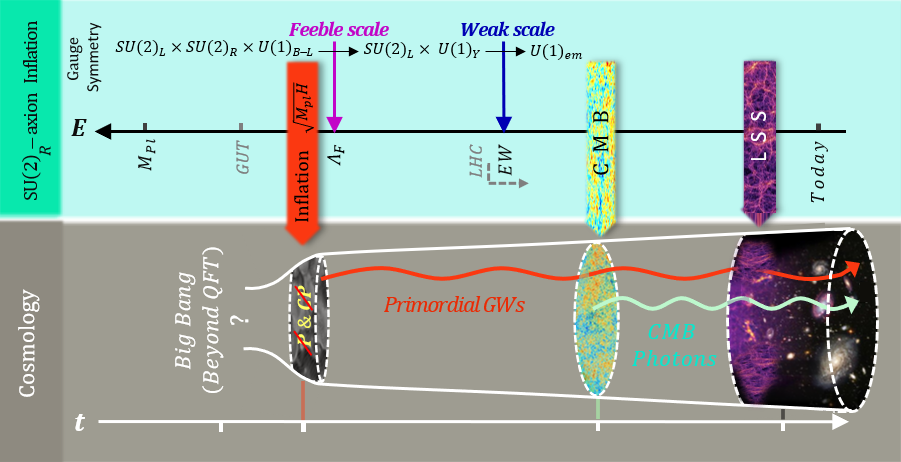}
    \caption{The $\rm{SU(2)_R}$-axion inflation throughout cosmic history. The seeds for CMB, LSS, DM, and BAU all share a common origin produced by quantum effects in inflation. }
    \label{fig:SU2R}
\end{figure}

$SU(2)_R$-axion inflation \cite{Maleknejad:2020yys} introduces a new framework for simultaneous baryogenesis and dark matter production, based on embedding axion-inflation in left-right symmetric models (LRSM) of particle physics. The source of asymmetry is spontaneous $CP$ violation in inflation, and the dark matter candidate is a right-handed neutrino.\footnote{One of the most well-studied leptogenesis scenarios is the
LRSMs \cite{Hati:2018tge, Dunsky:2020dhn}. The current proposal is an alternative mechanism that does not relay on the unconstrained
$CP$-violating phases in the neutrino sector.} The seeds for cosmic microwave background (CMB), large scale structure (LSS), dark matter (DM), and baryon asymmetry all share a common origin produced by quantum effects during inflation. Thus, this model can naturally explain the observed coincidences among cosmological parameters, i.e., $\eta_B = 0.3 P_{\zeta}$ and $\Omega_{DM}=5 \Omega_{b}$. This scenario has a distinct predictions on gravitational wave background (GWB) that could be detected by future CMB missions and GW detectors across 21 decades in frequencies \cite{Komatsu:2022nvu}. (See Fig. \ref{fig:SU2R})

Axion fields are abundant in string theory, well-motivated candidates
for the inflaton field \cite{Freese:1990rb, Pajer:2013fsa, McAllister:2014mpa} , and are naturally coupled to gauge fields.
Non-Abelian gauge fields may contribute to the physics of inflation while respecting the cosmological symmetries \cite{Maleknejad:2011sq, Maleknejad:2011jw, Adshead:2012kp}. It is often assumed that physics during inflation was $P$ and $CP$ symmetric. However, it is the perfect cosmic era for spontaneous $CP$ violation without the domain wall problem \cite{Melnikov1982CosmologicalCO, Dvali:2018txx}. From the phenomenological perspective, such inflation physics may be a natural setting for BAU \cite{Alexander:2004us, Maleknejad:2016dci, Caldwell:2017chz, Maleknejad:2020pec}. From the observational point of view, it predicts circularly polarized GWB and CMB parity-odd power spectra which can be used to probe such new physics  \cite{Lue:1998mq, Komatsu:2022nvu}. Originally proposed to explain $P$ violation in low energy processes \cite{Pati:1974y}, LRSM has a number of compelling consequences: e.g. natural $B-L$ symmetry \cite{Mohapatra:1980qe}, natural entailed seesaw mechanisms \cite{Mohapatra:1980yp}, and a solution to vacuum stability problem \cite{Maiezza:2016ybz}. 
By embedding axion-inflation in LRSM, we have two unknown energy scales, the scale of inflation  $\Lambda_{\rm{Inf}}$, and left-right symmetry breaking $\Lambda_F$. (See Fig. \ref{fig:SU2R})

\begin{figure}[h]
    \centering
    \includegraphics[width=0.9\textwidth]{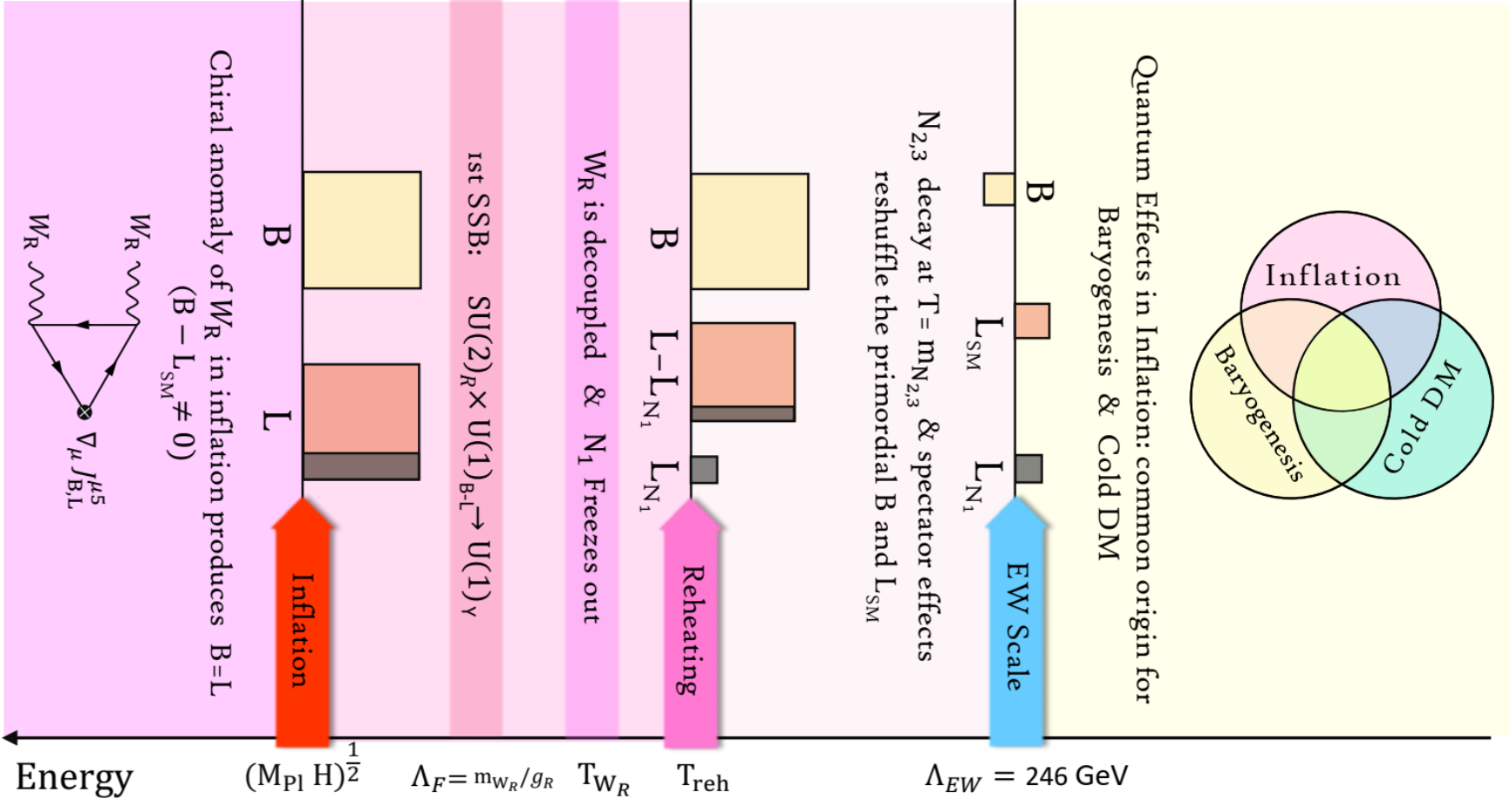}
    \caption{Summary of the new mechanism and evolution of baryons $B$ (yellow
box), SM leptons $L_{SM}$ (pink box) and RHNs ${\rm{L}}_N = \sum_{i=1}^3 {\rm{L}}_{N_i}$ (gray box).  Figure adopted from \cite{Maleknejad:2020pec}.}
    \label{fig:summary}
\end{figure}

The new setup proposed in \cite{Maleknejad:2020yys, Maleknejad:2020pec} extended the field content of the minimal LRSM with an axion field, which drives the cosmic inflation. It is assumed that the axion and $W_R$ gauge field are coupled by a Chern-Simons interaction.\footnote{We can couple the axion to both $SU(2)_R$ and $SU(2)_L$ gauge fields. However, the primordial left-handed fermions produced by $SU(2)_L$ will be completely washed out by the $SU(2)_L$ sphaleon effects \cite{Maleknejad:2020pec}.  } 
Both $P$ and $CP$ are spontaneously violated by the axion interactions. Without any interactions with the inflaton field, all massless gauge fields and fermions dilute exponentially during inflation due to conformal symmetry.  However, the axion-gauge field interaction breaks its conformal symmetry and generates $W_R$ gauge field in inflation. Besides, the right-handed gauge field is coupled to the right-handed fermions via the famous chiral anomaly  which creates right-handed chiral fermions coupled to it, i.e., SM baryons $B$, SM leptons $L_{\rm SM}$, and three Right-Handed Neutrinos (RHN) $L_{\rm N}$. By the end of inflation, equal (right-handed) baryon and lepton numbers are created in inflation, i.e. $B=L$. However, since $L\equiv L_{\rm SM}+L_{\rm N}$, there is a net $B-L$ in the SM sector. Shortly after inflation, the first spontaneous symmetry breaking (SSB) happens at $\Lambda_F$, and after a while at temperature $T_{W_R}$, the $SU(2)_R$ interactions freeze out. The lightest of RHNs with feeble Yukawa couplings (our DM candidate) is decoupled at this point, while $\rm{N}_{2,3}$ decay at $T=m_{\rm N_{2,3}}$. Between reheating and EW scale, the spectator effects reshuffle the primordial densities. The desired baryon to photon ratio today is \bea
\eta_{b} \approx \frac{\alpha_{\rm{Inf}}}{3\varepsilon^{\frac34}} ~\bigg(\frac{H}{M_{Pl}}\bigg)^\frac{3}{2},
\eea
where $\alpha_{\rm{Inf}} \equiv B-L_{\rm SM} $ is the value of the generated baryon minus lepton numbers in the SM sector by the end of inflation and $\varepsilon$ is the efficiency of reheating. Setting $\eta_b \simeq 6 \times 10^{-10}$, it demands $\Lambda_F \approx \Lambda_{\rm{Inf}}$,
i.e. the LR SSB should coincide with the geometrical
transition that ends inflation. Interestingly, it prefers
Left-Right symmetry breaking scales above $10^{10}~$GeV , which is in the range suggested by the non-supersymmetric SO(10) GUT with an intermediate left-right symmetry scale. Finally, if the lightest RHN is stable enough to make a dark matter candidate and makes all the DM today, its mass should be 
$m_{N_1} \approx 1 ~$GeV. The summary of this new mechanism  is presented in Fig. \ref{fig:summary}.

\subsection{QCD Baryogenesis}
\subsection*{Contributor: Seyda Ipek}

In the SM, the QCD sector is not thought to play a role in generating the BAU. There is no baryon violation in strong interactions and the interaction rates are too high to fall out of equilibrium. Furthermore the QCD phase transition from quark-gluon plasma to the hadronic phase is expected to be a crossover. On the other hand, there could have been large $CP$ violation in the QCD sector in the early universe, which necessitates an axion field to explain the lack of neutron electric dipole moment (EDM) at zero temperature. 
 
The QCD transition is described by the energy scale $\Lambda_{\rm QCD}$ when the QCD coupling constant becomes large. In \cite{Ipek:2018lhm} it was shown that this scale can be promoted into a dynamical quantity,
\begin{align}
    \Lambda(\langle \phi\rangle)=\Lambda_0 \exp\left(\frac{24\pi^2}{2n_f -33}\frac{\langle \phi\rangle}{M_*}\right)\,,
\end{align}
by introducing a scalar field $\phi$ that couples to the gluon field strength through the following BSM interactions
\begin{align}
    \mathcal{L}\supset -\frac14 \left(\frac{1}{g_{s0}^2}-\frac{\phi}{M_\ast}\right)G^{\mu\nu}G_{\mu\nu}\,.
\end{align}
Thus, the QCD scale can be made different by tuning the ratio of the scalar vev $\langle \phi \rangle$ to the new physics scale $M_\ast$.

\begin{figure}[t]
    \centering
    \includegraphics[width=\textwidth]{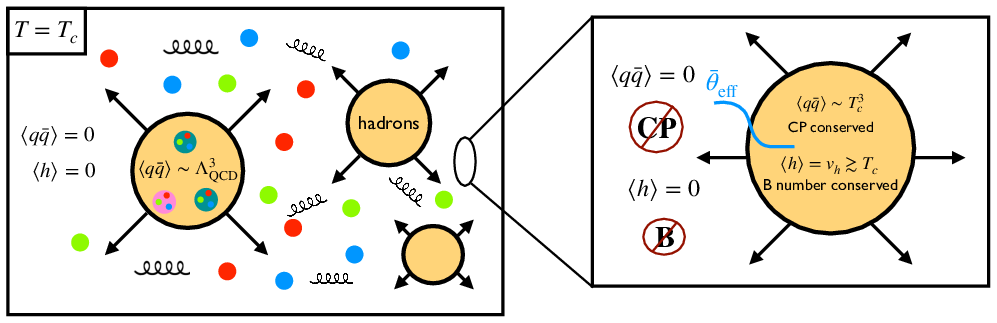}
    \caption{Schematic description of the baryon asymmetry production via the new QCD confinement scenario.}
    \label{fig:QCDBG}
\end{figure}

The value of $\langle\phi\rangle/M_\ast \sim -0.5$ is particulary interesting because then QCD sector confines at a temperature of hundreds of GeV, before the EW symmetry breaking. In this case all six quarks would be massless at the time of the QCD phase transition, which is expected to result in a first-order phase transition. The sphalerons would still be active right before QCD phase transition occurs. However, QCD transition triggers EW symmetry breaking due to chiral symmetry breaking. Hence, baryon number violation is turned off inside the bubbles. If we also add an axion field as a solution to the strong $CP$ problem to this scenario, we get a novel baryogenesis scenario. (A schematic description is shown in Figure \ref{fig:QCDBG}.) The baryon asymmetry produced through this mechanism is 
\begin{align}
    \eta \simeq 10^{-11}\sin\bar{\theta} \left(\frac{v_h}{\Lambda_{\rm QCD}}\right)\left(\frac{T_{\rm sph}}{T_{\rm reh}}\right)^3\,,
\end{align}
where $T_{\rm sph}$ is the temperature associated with the sphaleron processes while $T_{\rm reh}$ is the reheat temperature after the transition completes. Here $v_h$ is the Higgs VEV triggered by the QCD transition at $\Lambda_{\rm QCD}$. In the benchmark scenarios discussed in \cite{Croon:2019ugf}, $v_h/\Lambda_{\rm qCD}=1-4$. 

In \cite{Ellis:2019flb} this mechanism was extended to include variations of $SU(2)_L$ and $U(1)_Y$ from their SM values in the early universe. In this case, a change in $\alpha_2$ can be employed to turn the EW transition into a first-order phase transition and a change in $\alpha_3$ can be utilized to generate enough $CP$ violation.  

\subsection{Wash-in Leptogenesis and Leptoflavorgenesis}
\subsection*{Contributor: Kai Schmitz}

In standard thermal leptogenesis, $CP$ invariance and baryon-minus-lepton number $B\!-\!L$ are violated simultaneously during the decay of heavy right-handed neutrinos (RHNs).
This is, however, not a necessary condition for the successful generation of the BAU; the energy scales of $CP$ and $B\!-\!L$ violation may in fact be well separated from each other.
Imagine, \textit{e.g.}, that a new $CP$-violating out-of-equilibrium process, \textit{chargegenesis}, first creates one or several charge asymmetries $C_i$ in the SM plasma at very high temperatures.
In the type-I seesaw model, RHN interactions at much lower temperatures are then able to reprocess these primordial charges into a $B\!-\!L$ asymmetry, which is eventually converted to a baryon asymmetry by SM sphaleron processes.
In this scenario, known as \textit{wash-in leptogenesis}~\cite{Domcke:2020quw}, RHN interactions play the role of spectator processes that drive the thermal bath towards a new $B\!-\!L$-violating chemical equilibrium\,---\,they hence \textit{wash in} a $B\!-\!L$ asymmetry.

Wash-in leptogenesis represents a natural extension of the chemical transport in the SM thermal bath at high temperatures:
In standard scenarios, sphalerons translate from $B-L$ to $B$ after leptogenesis; wash-in leptogenesis now takes this idea a step further and translate from $C_i$ to $B\!-\!L$ after chargegenesis.
This significantly relaxes the requirements imposed by the Sakharov conditions and opens up a broad window for constructing new models, especially, in view of the many conserved SM charges $C_i$ at high temperatures~\cite{Domcke:2020kcp},
\begin{equation}
\label{eq:charges}
\mu_{C_i} \:\in\: \left\{\mu_u,\: \mu_B,\: \mu_{d-b},\: \mu_\tau,\: \mu_{u-c},\: \mu_\mu,\: \mu_{B_1-B_2},\: \mu_{d-s},\: \mu_{u-d},\: \mu_{2B_1-B_2-B_3},\: \mu_e \right\} \,.
\end{equation}
At $T \sim 10^{13\cdots15}\,\textrm{GeV}$, all of these chemical potentials are conserved; at $T \sim 10^{5\cdots6}\,\textrm{GeV}$, at least the chemical potential of right-handed electrons, $\mu_e$, is a conserved quantity.
In the framework of wash-in leptogenesis, it is thus possible to construct models of \textit{electrogenesis}, \textit{muogenesis}, etc.
It is not necessary to produce a $B$ or $B-L$ asymmetry right away.

Any process that can give rise to at least one of the charges listed in Eq.~\eqref{eq:charges} qualifies as a possible chargegenesis mechanism.
Important examples include (i) GUT baryogenesis, which produces $\mu_B$ among other charges and which is hence revived in the context of wash-in leptogenesis, and (ii) axion inflation coupling to the SM hypercharge gauge field~\cite{Anber:2015yca,Jimenez:2017cdr}, which populates the charges $\mu_u$, $\mu_B$, $\mu_\tau$, $\mu_\mu$, $\mu_{u-d}$, and $\mu_e$~\cite{Domcke:2019mnd}.
Both processes notably do not violate $B\!-\!L$ and only lead to a primordial $B\!+\!L$ asymmetry.
In order to understand the action of RHN interactions on the primordial charges after axion inflation, consider, \textit{e.g.}, nondegenerate RHN masses $M_i \sim 10^{5\cdots6}\,\textrm{GeV}$, such that, at $T \sim M_i$, the combination of all SM interactions except for the electron Yukawa interaction results in~\cite{Domcke:2020quw}
\begin{equation}
\label{eq:washin}
\begin{pmatrix}
		\mu_{\ell_e}    + \mu_\phi \\
		\mu_{\ell_\mu}  + \mu_\phi \\
		\mu_{\ell_\tau} + \mu_\phi
	\end{pmatrix}
	= 
	\begin{pmatrix}
		-\frac{5}{13} \\
		 \frac{4}{37} \\
		 \frac{4}{37}
	\end{pmatrix} \,\mu_e 
	-
	\begin{pmatrix}
		\frac{6}{13} &  0              & 0             \\
		0            &  \frac{41}{111} & \frac{4}{111} \\
		0            &  \frac{4}{111}  & \frac{41}{111} 
	\end{pmatrix}
	\begin{pmatrix}
		\mu_{\Delta_e}\\
		\mu_{\Delta_\mu}\\
		\mu_{\Delta_\tau}
	\end{pmatrix} \,,
\end{equation}
where $\Delta_\alpha = B/3 - L_\alpha$ ($\alpha = e,\mu,\tau$).
RHN decays and inverse decays at $T \sim M_i$ then impose the conditions $\mu_{\ell_\alpha} + \mu_\phi \approx \mu_{N_i} \approx 0$, which allows one to solve Eq.~\eqref{eq:washin} for the three flavored $B-L$ charges, $\sum_\alpha \mu_{\Delta_\alpha} = -3/10\,\mu_e$, where $\mu_e$ is related to the efficiency of gauge-field production during axion inflation~\cite{Gorbar:2021rlt,Gorbar:2021zlr}.
While this is only a minimal example for one specific chargegenesis mechanism and one specific RHN mass range, it already illustrates three important properties of wash-in leptogenesis: It is (i) independent of $CP$ violation in the RHN sector, (ii) can operate at RHN masses as low as 100 TeV, and (iii) is especially efficient in what is otherwise known as the strong-washout regime.

The observational signatures of wash-in leptogenesis depend on the source of high-scale $CP$ violation. 
Axion inflation, \textit{e.g.}, gives rise to rich phenomenology ranging from intergalactic magnetic fields over primordial black holes to gravitational waves, while GUT baryogenesis may be probed via classic GUT observables such as proton decay.
Finally, we also comment on an interesting variation of wash-in leptogenesis that does not proceed via the wash-in sequence $C_i \overset{\textrm{\tiny RHN}}{\longrightarrow} B\!-\!L \overset{\textrm{\tiny sph}}{\longrightarrow} B$, but instead via the sequence $C_i \overset{\textrm{\tiny CLFV}}{\longrightarrow} \Delta_\alpha \overset{\textrm{\tiny sph}}{\longrightarrow} B$ with $B-L = 0$ at all times and across all sectors.
This scenario is known as (wash-in) \textit{leptoflavorgenesis}~\cite{Mukaida:2021sgv} and relies on charged-lepton flavor violation (CLFV) to wash in nonzero $\Delta_\alpha$ charges that are subsequently converted to a baryon asymmetry by higher-order corrections to the sphaleron conversion formula~\cite{Khlebnikov:1988sr,Laine:1999wv}. 
This scenario can be tested in searches for CLFV processes such as $\mu \rightarrow e \gamma$ and $\mu \rightarrow e a$, where $a$ is an axion.

\subsection{Hylogenesis}
\subsection*{Contributors: Nikita Blinov, Hooman Davoudiasl, David E. Morrissey}

Hylogenesis\footnote{From {\it hyle}, which means ``matter" in Greek.} is a mechanism to generate the relic baryon and dark matter densities simultaneously~\cite{Davoudiasl:2010am,Davoudiasl:2011fj}. In this mechanism, global baryon number is extended such that one or more dark matter states are charged under the generalized group. Non-equilibrium $C$ and $CP$-violating processes in the early universe separate a net zero generalized baryon charge into a positive component consisting of baryons and a negative component made up of dark matter. Subsequent annihilation processes then deplete the symmetric components of the baryon and dark matter densities leaving behind only the equal and opposite asymmetries. A novel signature of hylogenesis is induced nucleon decay, in which dark matter destroys a stable nucleon through scattering.

In the minimal realization of Hylogenesis presented in Ref.~\cite{Davoudiasl:2010am}, the new states giving rise to the baryogenesis dynamics consist of (at least) two heavy Dirac fermions $X_a$ together with a lighter fermion $Y$ and complex scalar $\Phi$. Their interactions are taken to be
\bea
    -\mathscr{L} \ \supset \ \frac{\lambda_a}{M^2}\,X_a\,U^cD^cD^c + y_a\,X_a\,Y\,\Phi \ . 
    \label{eq:hylo1}
\eea
These couplings allow for a generalized $U(1)$ baryon number under which $[X] = 1$ and $[Y]+[\Phi] = -1$. In addition, the $Y$ and $\Phi$ states are assumed to have equal and opposite charges under a new dark $U(1)_x$ gauge invariance.

Hylogenesis begins with the (unspecified) non-thermal production of equal numbers of $X_1$ and anti-$X_1$ particles in the early universe. With the interactions of Eq.~\eqref{eq:hylo1}, the $X_a$ will then decay to quarks as well as the $Y$ and $\Phi$ particles. If these decays involve $C$ and $CP$ violation, the branching fractions of $X_a$ into quarks can be slightly different from those of the anti-$X_a$ into anti-quarks. To maintain CPT invariance, these differences must be compensated for with corresponding differences in the branching ratios  to the $Y$ and $\Phi$ states. Taken together, the net effect is a separation of the generalized baryon charge into quarks on the one hand and $Y$ and $\Phi$ on the other.

To preserve the baryon asymmetry, the asymmetry in the $Y$ and $\Phi$ must be (approximately) stable. This requires  
\begin{equation}
    |m_Y - m_\Phi| < m_p + m_e < m_Y + m_\Phi\,,
    \label{eq:stability}
\end{equation}
where $m_Y$, $m_\Phi$, $m_p$, and $m_e$ are the masses of $Y, \Phi$, the proton, and the electron, respectively.  One can show that 2~GeV$\lesssim m_Y, m_\Phi \lesssim 3$~GeV in the above minimal Hylogenesis framework~\cite{Davoudiasl:2010am}.  

An interesting aspect of Hylogenesis is that scattering processes can result in the exchange of baryon number between the visible and the dark sectors, leaving net baryon number the same~\cite{Davoudiasl:2010am,Davoudiasl:2011fj}.  In particular, a dark matter state from the galactic halo can transmute into a dark baryon by scattering from a nucleon and destroying it.  For example, this can happen via $p \,\Phi \to \pi^+\, \bar Y$, with $[Y]=-1$ and $[\Phi]=0$. Since the final state dark matter particle is invisible, this effectively mimics proton decay into a meson and a neutrino, $p\to \pi^+ \nu$, and was termed Induced Nucleon Decay~(IND).  

One can use chiral perturbation theory to get an estimate of the expected rate for IND scattering processes~\cite{Davoudiasl:2011fj}.  Remarkably, it turns out that if the UV scale $M$ and mass of $X$ are $\mathcal{ O}({\rm TeV})$ in Eq.~\ref{eq:hylo1}, the effective lifetime of nucleon IND processes in Earth's galactic neighborhood $\tau_{\rm IND}^{\oplus} \sim 10^{32}$~yr, close to those for standard nucleon decay in grand unified models.  Hence, IND signals of hylogenesis are potentially within reach of experiments that search for nucleon decay.  However, the kinematics of IND processes are different.  In particular, when the final state dark matter particle is lighter than that in the initial state, {\it i.e.} in down-scattering processes, the meson momentum can be typically ${\cal O}({\rm GeV})$, larger than meson momentum in standard nucleon decay counterparts.  Thus, for more accurate estimates of the effective nucleon lifetime in the presence of IND, non-perturbative lattice calculations are required (see, {\it e.g.}, Ref.~\cite{Aoki:2017puj}).

Different IND processes can arise from Eq.~\ref{eq:hylo1}, or from other implementations of the hylogenesis mechanism. For example, Ref.~\cite{Demidov:2015bea} studied processes like $\Phi n \to \bar{Y} \gamma$, $\Phi n \to \bar Y \gamma^*(e^+ e^-)$ and IND processes with multiple mesons in the final state. These reactions make use of additional couplings (e.g., neutron dipole moment or the electromagnetic interaction) which suppresses the rate compared to the minimal IND $p \,\Phi \to \pi^+\, \bar Y$ for a given $M$; however, they might be more easily observable with standard searches for nucleon decay. Ref.~\cite{Huang:2013xfa} instead examined a realization where the leading IND-like processes include $\Phi n \to \Phi^\dagger \bar\nu$, which can be probed through neutrino flux measurements from the Sun.

The hadronic operator in Eq.~\ref{eq:hylo1} can also be tested at $pp$ and $p\bar p$ colliders, where it mediates processes like $q q \to \bar q + X$, with on- or off-shell $X$ decaying to DM~\cite{Davoudiasl:2011fj,Demidov:2014mda}. This gives rise to a missing energy plus jets signature, which can enables the LHC to probe $X$ masses at the TeV scale (depending on the UV scale $M$). Further collider signatures can arise if hylogenesis emerges from a supersymmetric hidden sector~\cite{Blinov:2012hq}.

\subsection{Darkogenesis}
\subsection*{Contributor: Robert McGehee}

Darkogenesis or dark baryogenesis broadly refers to any mechanism in which an asymmetry is first generated in the dark sector before being transferred to the SM baryon asymmetry. Many models of Darkogenesis, including the original~\cite{Shelton:2010ta}, mimic models of electroweak baryogenesis while avoiding their pitfalls (\emph{e.g.} too large electric dipole moments \cite{ACME:2018yjb}). Their dark sectors often have dark sphalerons, contain large $CP$ violation and dark Higgs(es), and go through a strongly first order phase transition. Some implementations achieve the baryon asymmetry with minimal dark sectors~\cite{Hall:2019ank}, while many others additionally generate asymmetric dark matter and relate its abundance to the SM baryon's~\cite{Dutta:2006pt,Shelton:2010ta,Dutta:2010va,Petraki:2011mv,Walker:2012ka,Hall:2019rld,Ritter:2021hgu}. In fact, Darkogenesis scenarios provide some of the most compelling explanations of the intriguing closeness of the dark and ordinary matter abundances, $\Omega_c \approx 5 \times \Omega_b$ \cite{ParticleDataGroup:2020ssz}. 

While Darkogenesis encompasses a great breadth of models, a few illustrative realizations highlight both the common mechanisms as well as interesting testable signals that accompany such dark sectors. In \cite{Hall:2019ank}, a model is proposed to mimic electroweak baryogenesis in the dark sector with the minimum of necessary ``ingredients.'' The dark sector contains a gauged $\text{SU}(2)$, two dark Higgs doublets, two dark lepton doublets, and two dark right-handed singlets. The two dark Higgs permit tree-level $CP$ violation in the dark potential; a component of one lepton doublet has a large Yukawa and plays the role of the SM top in normal electroweak baryogenesis; the other doublet cancels Witten's anomaly; and the two singlets establish a neutrino portal. The dark sector undergoes a strongly first order phase transition which, together with the dark sphalerons, generates dark asymmetries. The neutrino portal then transfers some of this generated asymmetry to the SM where it is converted to the baryon asymmetry prior to SM electroweak symmetry breaking. 

Even such a minimal model produces a variety of detectable signals. Exotic SM Z and Higgs decays to light singlets may be discovered at future $e^+ e^-$ Higgs factories \cite{Liu:2016zki}. The dark phase transition can generate gravitational waves within the reach of a variety of future observatories such as LISA \cite{Audley:2017drz}, LIGO \cite{KAGRA:2013rdx}, ET \cite{Hild:2010id}, and BBO \cite{Yagi:2011wg}, depending on the temperature of the phase transition. This iteration of Darkogenesis also predicts an irreducible contribution to the number of relativistic degrees of freedom in the early Universe which will be fully probed soon by future CMB stage 3 experiments \cite{SPT-3G:2014dbx,POLARBEAR:2015ixw,BICEP3:2016pqy,ACTPol:2016kmo} and stage 4 ones \cite{CMB-S4:2016ple}.

Expanding the dark sector into a full copy of the SM gauge group and one SM matter generation (including a single right-handed neutrino) can additionally generate asymmetric dark matter \cite{Hall:2019rld}. This in turn can lead to even more complementary signals, including visibly decaying dark photons detectable at future experiments \cite{Fabbrichesi:2020wbt} as well as detectable nuclear recoils in dark matter direct detection experiments \cite{XENON:2018voc,DARWIN:2016hyl}. Other models of Darkogenesis vary the gauge/particle content of the dark sector, the portal used for asymmetry transfer, and the scale of the dark-sector phase transition, but often follow an analogous story to the minimal model's. In addition, the dark-sector phase transition can be used to ``filter'' heavy dark matter in a CPV way to create a dark chiral asymmetry which is converted to a lepton and then baryon asymmetry~\cite{Baker:2021zsf}. 
It is also possible to use CPV decays of a parent particle in the dark sector (as in leptogenesis) to source the initial asymmetry~\cite{Haba:2010bm}.

\subsection{WIMP-Triggered Baryogenesis}
\subsection*{Contributors: Yanou Cui, Arnab Dasgupta, Michael Shamma and Brian Shuve}

WIMP-triggered baryogenesis models are inspired by recognizing that thermal freeze-out of WIMPs can provide the Sakharov out-of-equilibrium condition necessary for successful baryogenesis. The two main schools of producing the baryon asymmetry with WIMPs include: through their annihilation around the freezeout time (e.g. \cite{Cui:2011ab,Davidson:2012fn,McDonald:2010toz,Dasgupta:2019lha}), and through the post-freezeout decays of a meta-stable WIMP (e.g. \cite{Cui:2012jh,Cui:2013bta,Cui:2020dly,Chu:2021qwk}). These scenarios also provide new possibilities addressing the coincidence $\Omega_\text{DM}\sim5 \Omega_\text{B}$ between DM and baryon abundance.\\

\noindent\textbf{WIMPy baryogenesis from WIMP dark matter annihilation.} This type of model is based on the subtle fact that during the WIMP freeze-out, the net DM annihilation or departure from equilibrium becomes significant at $T\sim m_{\rm DM}$, which occurs before the typical freezeout temperature $T_{\rm fo}\sim m_{\rm DM}/O(10)$. This time separation allows the build-up of baryon asymmetry through DM depletion before its freezeout, provided that the annihilation is $CP$- and $B$ (or $L$)-violating. Washout processes suppress the produced baryon asymmetry, but these processes can slow down to below the Hubble rate before $T_{\rm f0}$. Generically, upon solving the Boltzmann equations for the DM and baryon abundances one finds the following approximate relation for the co-moving baryon asymmetry
\begin{equation}\label{eq:wimpy}
Y_{\Delta B}(T\rightarrow0)\approx\frac{\epsilon}{2}\left[Y_\text{DM}(T_\text{washout})-Y_\text{DM}(T\rightarrow0)\right]
\end{equation}
where $Y_\text{DM}(T_\text{washout})$ is the co-moving DM density at the time of washout processes freezing out, $Y_\text{DM}(T\rightarrow\infty)$ is the observed co-moving DM density, and $\epsilon$ is the $CP$ asymmetry factor defined as the net baryon asymmetry obtained for each DM annihilation. 

A specific realization of WIMPy baryogenesis from annihilations is one in which WIMP annihilations violate lepton number and produce an asymmetry in leptons before being converted into the observed baryon asymmetry by electroweak sphalerons. A  model realizing this scenario includes the following interations: 
\begin{equation}\label{eq:leptowimpy}
\mathcal{L}\supset\left(\lambda_iX^2+\lambda^\prime_i\bar{X}^2\right)S_i+\lambda_{\psi_i}L\psi S_i+\mathrm{h.c.},
\end{equation}
where DM consists of a gauge-singlet pair of Dirac fermions $X$ and $\bar{X}$ coupled to pseudo-scalar gauge singlets $S_{1,2}$. Additionally, the pseudo-scalar singlets couple to weak-scale $SU(2)_L$ doublet fermions $\psi$ and the left-handed SM lepton doublet.  $X-\bar{X}$ annihilation to $L,\psi$ trigger baryogenesis. The requirement that washout processes become ineffective before $X$ freezeout can be realized for $m_\psi\gtrsim m_X$, while $m_\psi< 2m_X$ is necessary for $X$ annihilation to be kinematically allowed. For further details see \cite{Cui:2011ab}.\\ 

\noindent\textbf{WIMPy baryogenesis from unstable intermediate particles.}
Another interesting mechanism to create the BAU involves  the interference of two independent sub-amplitudes of an annihilation or a decay process \cite{Dasgupta:2019lha}. 

\begin{figure}[t!]
    \centering
\includegraphics[scale=1,trim={6.4cm 22.6cm 6.6cm 4cm},clip,width=0.4\textwidth]{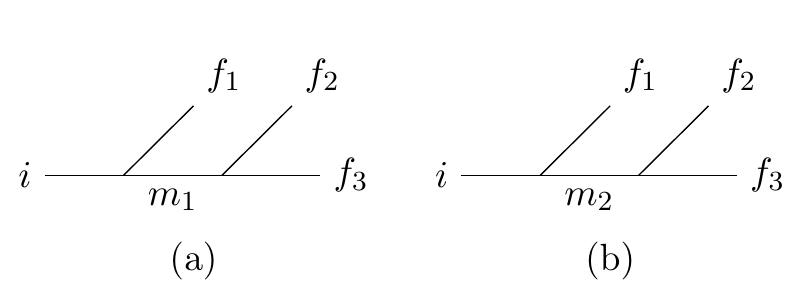}
~~~~~
\includegraphics[width=0.4\textwidth]{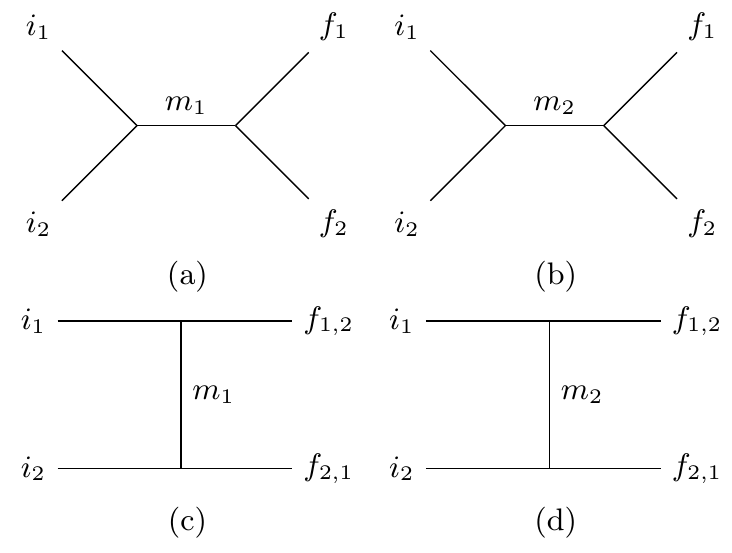}
\caption{Generic topologies for tree-level (left) $1\to 3$ and (right) $2\rightarrow2$ subprocesses that can give rise to a nonzero lepton or baryon asymmetry. Here $i_{1,2}$ and $f_{1,2}$ are respectively the initial and final states, and $m_{1,2}$ are the masses of two different mediators. }
\label{fig:woloops22}
\end{figure}

To realise such scenarios via $2\rightarrow 2$ scatterings or $1\rightarrow 3$ decays (see Fig. \ref{fig:woloops22}) which involves two \emph{different} intermediate state particles, with the outgoing particles (or decay products) carrying net nonzero baryon or lepton number. 

As an example we will consider the $2\rightarrow 2$ scattering case with the initial states $i_1,i_2$ and with only two sub-processes for the final states $f_1,f_2$ (here $i_{1,2}$ and $f_{1,2}$ generically stand bosons and/or fermions), mediated by intermediate-state particles of mass $m_1$ and $m_2$, respectively. Now, the total amplitude for this process is given as:
\begin{align}
    \mathcal{M} &= \left(\mathcal{C}_1\mathcal{M}_1 + \mathcal{C}_2\mathcal{M}_2\right)\mathcal{W}
\end{align}
where $\mathcal{C}_i$ contain only the couplings, $\mathcal{W}$ contains the wave functions for the incoming and outgoing particles and $\mathcal{M}_i$ stand for the rest of the subamplitudes. The corresponding amplitude for the conjugate process $\bar{i}_1\bar{i}_2\rightarrow \bar{f}_1\bar{f}_2$ is 
\begin{align}
    \bar{\mathcal{M}} &= \left(\mathcal{C}^*_1 \mathcal{M}_1 + \mathcal{C}^*_2\mathcal{M}_2 \right)\mathcal{W}^*.
\end{align}
And hence the asymmetry we get by comparing the modular squares of the amplitude as 
\begin{align}
    \delta &= -4\rm{Im}[\mathcal{C}_1\mathcal{C}^*_2]\rm{Im}[\mathcal{M}_1\mathcal{M}^*_2]|\mathcal{W}|^2,
    \label{eq:asymgen}
\end{align}
where $\rm{Im}[\mathcal{C}_1\mathcal{C}^*_2]$ is coming purely from the couplings, which is required to be non-zero for $CP$ violation, and $\rm{Im}[\mathcal{M}_1\mathcal{M}^*_2]$ incorporates the imaginary parts of the subamplitudes $\mathcal{M}_{1,2}$, which is reminiscent of the imaginary part coming from the interference of tree and loop-level diagrams in the $1\rightarrow 2$ decay scenario.
In  the $2\rightarrow 2$ process we will have the resummed propagator as 
\begin{align}
    \mathcal{M}_j &= \frac{A_j}{x_j - m^2_j + im_j\Gamma_j}
\end{align}
where $j=1,2$, $x_j=s,t,u$ are the Mandelstram variables, $m_j$ and $\Gamma_j$ are respectively the mediator masses and widths, and $A_j$ are some arbitrary real parameters. The imaginary parts from the product of  subamplitudes is given as 
\begin{align}
    \rm{Im}[\mathcal{M}_1\mathcal{M}^*_2] &= \frac{A_1A_2\left[(x_1-m^2_1)m_2\Gamma_2 - (x_2-m^2_2)m_1\Gamma_1\right]}{\left[(x_1-m^2_1)^2+m^2_1\Gamma^2_1\right]\left[(x_2-m^2_2)^2+m^2_2\Gamma^2_2\right]}
    \label{eq:asym22}
\end{align}
which is nonzero as long as the numerator is nonvanishing i.e $(x_1-m^2_1)m_2\Gamma_2\neq(x_2-m^2_2)m_1\Gamma_1$. There are two distinct possibilities for the tree-level $2\rightarrow 2$ case shown in Fig.\ref{fig:woloops22}:
\begin{enumerate}
    \item[(i)] If both of the subprocesses are  
    $s-$channel [cf. Fig.\ref{fig:woloops22}(a)+(b)], one just 
    needs to replace $x_{1,2}$ by $s$ in Eq.\eqref{eq:asym22}. 
    In this case, the $CP$-asymmetry factor $\delta$ in 
    Eq.\eqref{eq:asymgen} can be largely enhanced in the vicinity of resonance (s), with 
    $s-m^2_i\simeq m_i\Gamma_i$ (with $i=1,2$), similar to the enhancement effect in resonant leptogenesis~\cite{Pilaftsis:2003gt, Dev:2017wwc}.
    \item [(ii)] If one of the sub-amplitudes is in the $s$-channel and the other one in the $t$- or $u$-channel [cf.~Fig.~\ref{fig:woloops22} (a)+(d) or (b)+(c)],
one can safely take the imaginary part for the $t$- or $u$-channel propagator. For illustration, we take ${\cal M}_1$ as the $s$-channel and ${\cal M}_2$ as the $x$-channel ($x=t$ or $u$) amplitude. In this case, Eq.~\eqref{eq:asym22} can be simplified to
\begin{eqnarray}
{\rm Im} [{\cal M}_1 {\cal M}_2^\ast] \ \simeq \ -
\frac{A_1 A_2 m_1\Gamma_1}{\left[(s-m^2_1)^2 + m^2_1\Gamma^2_1\right](x-m^2_2)} \,,
\label{eq:6}
\end{eqnarray}
which is proportional to the $s$-channel mediator width $\Gamma_1$. Here, the $CP$-asymmetry could also be largely enhanced at the $s$-channel resonance, i.e. $s-m_1^2 \simeq m_1 \Gamma_1$.
\end{enumerate}

\noindent
\textbf{Baryogenesis triggered by metastable WIMP decay.}
In this type of model, the baryon asymmetry arises from the $CP$- and $B$($L$)-violating decays of a metastable WIMP parent $\chi_B$ after its annihilation freezes out. The late decay may occur in a wide time window prior to  BBN, while the predicted baryon asymmetry is not sensitive to the lifetime of $\chi_B$. Because the WIMP is long lived, washout processes are typically suppressed at the time of WIMP decay. As a result, the baryon asymmetry in this weak washout regime can be directly related to the density of $\chi_B$ at the time of its freeze-out:  
\begin{equation}
Y_\text{B}(0)\approx\epsilon Y_{\chi_B}(T_\text{f.o.}),~\Omega_\text{B}(0)=\epsilon\frac{m_p}{m_{\chi_B}}\Omega_{\chi_B}^{\tau\rightarrow\infty}
\end{equation}
where $\Omega_{\chi_B}^{\tau\rightarrow\infty}$ is the ``would-be" abundance of $\chi_B$ in the limit that it is stable, $m_p$ is the proton mass and $\epsilon$ is the $CP$ asymmetry factor. Therefore, a generalized WIMP miracle applies to baryon abundance. Note that this scenario provides a more direct connection between the baryon asymmetry and a DM-like abundance; in WIMPy baryogenesis, the asymmetry is determined by the DM density at washout decoupling, not the final DM abundance.

A minimal realization of this scenario is given by
\begin{equation}\label{eq:bg4wimpsL}
\Delta\mathcal{L}=\lambda_{ij}\phi d_id_j+\alpha_i\chi_B\bar{u}_i\phi+\beta_i\psi\bar{u}_i\phi+\eta\chi_B^2S+\gamma|H|^2S+\text{h.c.}
\end{equation}
where all couplings can be complex, $H$ is the SM Higgs boson; $d_i$ and $u_i$ are right- handed SM quarks with flavor indices $i = 1, 2, 3$; $\phi$ is a di-quark scalar with the same SM gauge charge as $u$; $\chi_B$ and $\psi$ are SM singlet Majorana fermions, and $S$ is a singlet scalar. The embedding of this mechanism in supersymmetry theories can be found in \cite{Cui:2012jh,Cui:2013bta}. 
Recent developments related to this scenario include, for instance, WIMP Cogenesis and Dark Freezeout Cogenesis.   \cite{Pierce:2019ozl,Cui:2020dly,Chu:2021qwk}. Additionally, there have been developments recently with WIMP-like models of baryogenesis from annihilation in \cite{Kumar:2013uca,Kang:2020ena,Chun:2020vxo,Ghosh:2020lma,Arakawa:2021wgz}
\vspace{0.1in}
\\
\textbf{Phenomenology.} The phenomenology predicted by WIMP baryogenesis mechanisms is rather rich, including various signals relevant for DM direct and indirect detection experiments, as well as precision frontier experiments (e.g. EDM measurements). Additionally, these models often contain new particles with masses and interactions at or near the electroweak scale, and thus can be within the reach of the current or near future particle collider experiments. Notably, in the case of WIMP baryogenesis from decays, the WIMP parent must survive its thermal freeze-out time in order to meet  out-of-equilibrium Sakharov condition, thus with a lifetime
\begin{equation}
\tau_\text{WIMP}\gtrsim\left(\frac{T_\text{f.o.}}{100~\text{GeV}}\right)~10^{-10}~\text{sec}
\end{equation}
This relatively long lifetime corresponds to a decay length, $l\sim1~\text{mm}$, which is intriguingly around the tracking resolution scale of detectors at collider experiments such as the LHC or future high luminosity experiments. As such, once the meta-stable WIMP is produced, its subsequent decays would generate displaced vertex signatures \cite{Cui:2014twa}. WIMP baryogenesis has become a benchmark case for long-lived particle searches at current/planned collider experiments \cite{ATLAS:2019qrr,Curtin:2018mvb,deBlas:2018mhx}.

\subsection{Gaugino Portal Baryogenesis}
\subsection*{Contributor: Bibhushan Shakya}

Supersymmetry remains one of the best motivated extensions of physics beyond the Standard Model. Supersymmetric constructions have been long known to provide viable baryogenesis mechanisms via R-parity violating (RPV) decays of heavy superpartners in the early Universe \cite{Dimopoulos:1987rk,Claudson:1983js,Rompineve:2013grm,Cui:2012jh,Cui:2013bta,Arcadi:2015ffa,Barbier:2004ez}. While LHC and direct detection measurements place severe constraints on weak scale supersymmetry \cite{Perelstein:2011tg,Amsel:2011at,Perelstein:2012qg}, in particular the Minimal Supersymmetic Standard Model (MSSM), supersymmetry might be realized at higher energies or in secluded sectors, with interesting phenomenological consequences \cite{Barnes:2020vsc,Barnes:2021bsn}. Such frameworks provide a novel and natural possibility for baryogenesis, via portal interactions between the visible and hidden sectors, where the portal interactions can naturally lead to favorable conditions for low scale baryogenesis. 

\begin{figure}[t!]
    \centering
    \includegraphics[width=0.8\textwidth]{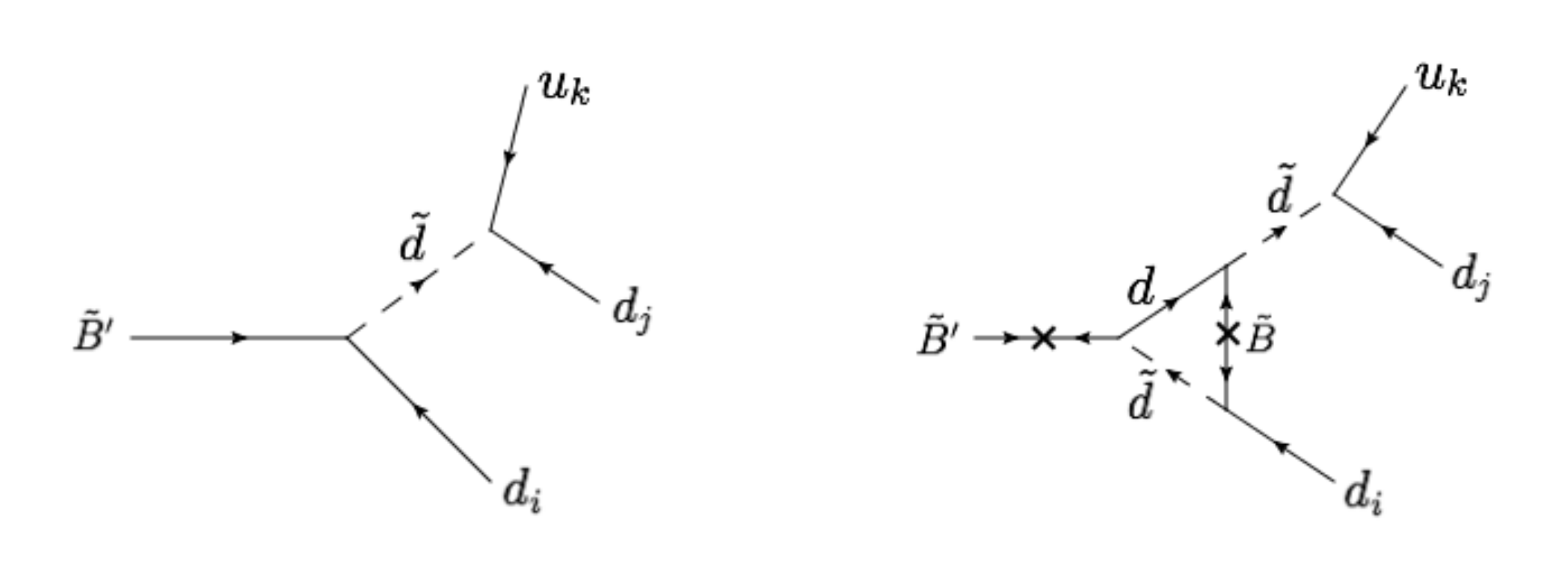}
    \caption{Interference between tree and loop decays of the hidden gaugino $\tilde{B}$ that produces a baryon asymmetry. In this instance, the nonvanishing $CP$ phase resides in the product of the two gaugino masses, represented by mass insertions on the loop diagram.}
    \label{fig:decay}
\end{figure}

A specific realization of this idea is gaugino portal baryogenesis \cite{Pierce:2019ozl}, which makes use of gaugino mixing between a hidden sector gaugino $\tilde{B}'$ and the bino $\tilde{B}$ of the MSSM, which is the supersymmetric counterpart of the familiar and extensively studied gauge kinetic mixing portal. The BAU is populated through late decays of the $\tilde{B}'$, which are produced via either freeze-in or freezeout processes in the early Universe. The Sakharov conditions are readily satisfied: baryon number violation is realized through R-parity violating decays of the $\tilde{B}'$; $CP$ violation arises from the interference of tree and loop level decays of the $\tilde{B}'$, with a nonvanishing $CP$ phase (in this case) in the gaugino masses (see Fig.\ref{fig:decay}), and the out-of-equilibrium condition is provided by late decays of the $\tilde{B}'$.

The small portal coupling $\epsilon$ between the two sectors serves two main purposes in the context of baryogenesis (see \cite{Pierce:2019ozl} for details). First, it naturally gives rise to a small decay width for the hidden gaugino $\tilde{B}'$, crucial to avoid the washout of the produced baryon asymmetry, without any large mass hierarchies or unnaturally small parameters, and without additionally suppressing the fraction of $\tilde{B}'$ decays that produce an asymmetry. Second, while large $CP$ violation involving particles below the PeV scale is strongly constrained by electric dipole moment (EDM) measurements \cite{Altmannshofer:2013lfa,Cesarotti:2018huy}, the portal coupling naturally allows the means to circumvent this constraint by realizing $CP$ violation in the hidden sector, so that its contribution to EDMs is $\epsilon^2$ suppressed. As a result, baryogenesis can be realized in the above setup for $\tilde{B}'$ masses as low as $10$ GeV (and as high as $10^8$ GeV). 

For not-too-small values of $\epsilon$, the hidden gaugino $\tilde{B}'$ can be produced at colliders and be observed as a long-lived particle with RPV decays. Depending on the nature of the RPV coupling, complementary signals can also be seen in low energy experiments, such as dinucleon decays, or neutron-antineutron oscillation if the coupling involves first generation quarks \cite{Grojean:2018fus}.

\subsection{Freeze-In Baryogenesis via Dark Matter Oscillations}
\subsection*{Contributors: Brian Shuve and David Tucker-Smith}

In models of freeze-in DM \cite{McDonald:2001vt,Hall:2009bx,Bernal:2017kxu}, the DM states are out of equilibrium for most or all of cosmic history. Thus, freeze-in DM satisfies the out-of-equilibrium Sakharov condition and can be responsible for baryogenesis \cite{Hall:2010jx,Hook:2011tk,Unwin:2014poa,Shuve:2020evk,Goudelis:2021lra,Berman:2022oht}. In a recently proposed mechanism, the production, oscillation, and subsequent scattering of two DM mass eigenstates can be simultaneously responsible for  the DM abundance as well as baryogenesis \cite{Shuve:2020evk,Berman:2022oht}. Because baryogenesis occurs at higher order in the out-of-equilibrium coupling than DM production, there is a tension between obtaining a sufficiently large baryon asymmetry without getting too much DM. The result is that there exists a bounded, potentially testable, region of parameter space for viable DM and baryogenesis.

The mechanism functions similarly to ARS leptogenesis \cite{Akhmedov:1998qx,Asaka:2005pn}, which has right-handed neutrinos as the oscillating, out-of-equilibrium states. Here, we discuss a concrete scenario where there exists one or more BSM scalars, $\Phi$, which are electroweak singlets with hypercharge $-1$ \cite{Berman:2022oht}. They interact with the DM mass eigenstates, $\chi_i$, and SM leptons via
\begin{equation}\label{eq:lepto_osc_lag}
\mathcal{L} \supset - F_{\alpha i}\,e_\alpha ^{\rm c}\Phi \chi_i + \mathrm{h.c.}.
\end{equation}
Note that a similar mechanism functions if $\Phi$ carries QCD charge and couples to quarks; see \cite{Shuve:2020evk}. 

There exist three identified classes of models for freeze-in leptogenesis via DM oscillations. {\bf (1) Minimal Model:}~Eq.~\eqref{eq:lepto_osc_lag} represents the entire model content and interactions of the theory. In this case, asymmetries in individual SM lepton flavors are generated at $\mathcal{O}(F^4)$, but the total lepton asymmetry arises only at $\mathcal{O}(F^6)$ due to flavor-dependent washout effects. Because the total asymmetry is generated at such a high order, the scenario requires $M_\Phi\lesssim1.5$ GeV, allowing for a longer period of DM oscillations and making the new scalar typically within reach of colliders. Additionally, the Minimal Model favors a nearly massless lightest DM ($\chi_1)$ state, which contributes to dark radiation, and a heavier DM eigenstate $\chi_2$ with mass $\lesssim100$ keV (constituting warm DM for much of the parameter space). For typical parameters, $\chi_1$ comes into equilibrium, leading to a breakdown of the perturbative expansion and mitigating the higher-order asymmetry compared to DM production. The larger couplings in this scenario lead to a prompt decay of $\Phi\to\ell\chi$ at colliders for much of the viable parameter space. {\bf (2) UVDM}:~in this model, there exist two scalars, $\Phi_1$ and $\Phi_2$, and consequently two sets of Yukawa couplings, $F_{\alpha i}^1$ and $F_{\alpha i}^2$. In this case, a total lepton asymmetry arises at $\mathcal{O}[(F^1)^2(F^2)^2]$, yielding a larger asymmetry for a given DM density. The lightest scalar can be much heavier (up to $\sim20$ TeV), although for generic Yukawa textures  the lightest $\Phi$ must be at the TeV scale. Additionally, a broader range of DM masses is allowed, with $M_{\chi 2}\lesssim$ MeV and $M_{\chi_1}$ heavy enough that it is not necessarily hot or warm. Because of the smaller couplings needed for baryogenesis, the lightest scalar $\Phi$ can be either prompt or displaced at colliders. {\bf (3) Z2V:} In the first two models, there exists a $Z_2$ symmetry that stabilizes DM. There  are, however, additional allowed interactions of $\Phi$ if the $Z_2$ symmetry is broken:~a neutrino portal coupling $h_{\alpha i} \ell_\alpha H \chi_i$ and a scalar coupling $\lambda_{\alpha\beta} \ell_\alpha \tilde\ell_\beta \Phi^*$. Because of X-ray line constraints, $h_{\alpha i}$ has to be so small as to play no role in leptogenesis. However, the coupling $\lambda_{\alpha\beta}$ essentially assigns a lepton number of $-2$ to $\Phi$ if the coupling is in equilibrium, and this allows the generation of a total lepton asymmetry at $\mathcal{O}(F^4\lambda^2)$ (or, $F^4$ if $\lambda$ comes into equilibrium) even with a single scalar. Like the UVDM model, $\Phi$ is often at the TeV scale but can be $\gtrsim$ 10 TeV for $\chi_1$ acting as dark radiation, and $M_{\chi_2}$ can be as heavy as the MeV scale. When the Z2V coupling is sufficiently large to be in equilibrium at the time of leptogenesis, $\Phi$  decays promptly.

The mechanism of freeze-in baryogenesis via DM oscillations naturally points towards several interesting energy scales:~TeV for the $\Phi$ mass, $10$ keV for the heavier DM mass eigenstate, and decay lengths up to the cm scale or longer. The model  therefore predicts signals that can be tested with a wide variety of probes \cite{Shuve:2020evk,Berman:2022oht}. Searches for SM fermion + missing momentum at colliders is a powerful probe, although in the case of leptogenesis we emphasize that $\Phi$ can decay to multiple flavors of leptons, weakening existing searches for sleptons. The model also has implications for structure formation and dark radiation, and in the Z2V model can also give rise to a number of other phenomena such as X-ray lines from DM decay, flavor violation, and contributions to $(g-2)_\mu$.

\subsection{Baryogenesis Through Particle-Antiparticle Oscillations}
\subsection*{Contributor: Seyda Ipek}

Pseudo-Dirac fermions have both Dirac and Majorana masses. The mass eigenstates are a mixture of particle and antiparticle interaction states and thus they can undergo particle--antiparticle oscillations similar to neutral meson oscillations. Furthermore, if both the particle and the antiparticle are allowed to decay into the same final state, there can be $CP$ violation due to a physical phase difference in the respective coupling constants~\cite{Ipek:2014moa}. The oscillations can enhance $CP$ violation in the parameter regime where the mass difference between the heavy and light mass eigenstates ($\Delta m$) is the same order of magnitude as the decay width of the particles ($\Gamma$). Namely, there can be a large amount of $CP$ violation if the particle/antiparticle system oscillates a few times before they decay. Conversely, if the oscillations are too fast ($\Delta m \gg \Gamma$), the $CP$ violation is washed out and if they are too slow ($\Delta m \ll \Gamma$) decay happens before oscillations and $CP$ violation is again reduced. 

In the early universe, the oscillation dynamics are affected by both the expansion of the universe and the interactions of the pseudo-Dirac fermions with the SM plasma. For example, oscillations do not start until the Hubble rate drops below the oscillation frequency, $H(T)<\omega_{\rm osc}=\Delta m$. For a mass difference of $\sim 10^{-4}~$eV and smaller, this means that oscillations would be delayed until the temperature of the universe drops below the mass of the particles, $T<M$. Then, in order for the $CP$ violation in these oscillations to be important, the decays should also be delayed, causing the particles to decay out of equilibrium. Assuming these decays are baryon-number (or lepton-number) violating, all three of the Sakharov conditions could be satisfied. 

The interactions of the pseudo-Dirac fermions with the SM plasma can further hinder the oscillations in the early universe. These interactions can either be \emph{flavor-blind} or \emph{flavor-sensitive} depending on if the Lagrangian is symmetric or antisymmetric under $\psi \to \psi^c$, respectively. Here $\psi$ is the particle state and $\psi^c$ is the antiparticle state. 

The Boltzmann equations that govern the time evolution of the density matrix $Y\equiv n/s \propto \sum_{\psi,\psi^c}|\psi_i\rangle\langle\psi_i|$ for particles $\psi$ and antiparticles $\psi^c$ are given as
\begin{align}
    z H \frac{d\mathbf{Y}}{dz} = -i (\mathbf{H}\mathbf{Y}-\mathbf{Y}\mathbf{H}^\dagger) -\frac12\sum\limits_{+,-}\Gamma_{\pm} [O_\pm,[O_\pm,\mathbf{Y}]]-\sum\limits_{+,-}s\langle \sigma v\rangle_\pm \left( \frac12 \{\mathbf{Y},O_\pm\overline{\mathbf{Y}}O_\pm\}-Y_{\rm eq}^2 \right)\,,
\end{align}
where $z=M/T$, $H$ is the Hubble rate and $s$ is the entropy density. The first term on the left-hand side describes the oscillations with the Hamiltonian $\mathbf{H}=\mathbf{M}-i\mathbf{\Gamma}$.  $\Gamma_+/\Gamma_-$ is the rate of inelastic scatterings for flavor-sensitive/blind interactions respectively, $\langle \sigma v\rangle$ is the thermally-averaged annihilation cross section and $O_\pm = {\rm diag}(1,\pm 1)$ differentiates between flavor-sensitive and flavor-blind interactions. Note that the second term is zero for flavor-blind interactions. 

These equations can be solved numerically for a model-independent, generic scenario~\citep{Ipek:2016bpf}. The resulting baryon asymmetry can be well approximated by 
\begin{align}
\Delta_B\simeq \epsilon \Sigma_\psi(z_{\rm osc})~,
\end{align}
where $z_{\rm osc}$ is when oscillations start and $\Sigma_\psi = Y_\psi + Y_{\psi^c}$. The $CP$-violation is quantified by the parameter $\epsilon=(\Gamma(\psi \to BX)-\Gamma(\psi\to\bar{B}X))/\Gamma$, where $X$ is a state with zero baryon number. 

This mechanism can be easily realized in $U(1)_R$-symmetric MSSM (MRSSM), see \emph{e.g.} \cite{Randall:1998uk}. In MRSSM, gauginos are necessarily pseudo-Dirac fermions. If there are also R-parity-violating interactions, then a pseudo-Dirac bino can go under $CP$-violating oscillations and $B$-violating decays. This scenario is realized for representative mass scales of $\mathcal{O}(100~{\rm GeV})$ bino and $\mathcal{O}(10~{\rm TeV})$ sfermions.

\subsection{Mesino Oscillations and Baryogenesis}
\subsection*{Contributor: Akshay Ghalsasi}
In the most popular models of baryogenesis (Electroweak Baryogenesis, vanilla Leptogenesis etc.), the baryon asymmetry is produced at very high temperatures, above or around the weak scale. However the cosmology of most supersymmetric extensions of standard model prefer a low reheating scale ~\cite{Pagels:1981ke,Weinberg:1982zq,Ellis:1986zt,Ellis:1990nb,Moroi:1993mb,Kallosh:1999jj,Bolz:2000fu,Banks:1993en}. Models of light scalar dark matter also prefer low inflation scale to suppress isocurvature perturbations \cite{Banks:1993en,Linde:1984ti,Turner:1989vc,Linde:1991km,Lyth:1992tx,Fox:2004kb,Sikivie:2006ni}. Models of dynamic relaxation of the weak scale also require the universe to reheat below the weak scale \cite{Arkani-Hamed:2016rle}.  Low inflation or reheat scale implies a low scale  to achieve baryogenesis, even as low as the QCD scale (1-200 \MeV). Although models for low scale baryogenesis exist \cite{Claudson:1983js, Dimopoulos:1987rk}, they require the existence of long lived heavy particles, whose decay dilutes away any baryon asymmetry that it produces. Moreover, these models typically require large $CP$ violation to get the required baryon asymmetry, which suffers from constraints on electric dipole moments of particles.

In \cite{Ghalsasi:2015mxa} a model for low scale baryogenesis was proposed via mesino oscillations. A mesino is a bound state between $SU(3)_c$-charged heavy scalar and a standard model quark that forms after the QCD phase transition. At the QCD phase transition, an equal number of mesino and anti-mesino are created. Neutral mesinos can oscillate into anti-mesinos and vice-versa. Analogous to the neutral kaon system in SM, interference between on and off-shell oscillations gives rise to $CP$ violation preferring mesinos over anti-mesinos. Out of equilibrium, baryon number violating decays of the scalar in the mesino then gives us the required baryon asymmetry.

The model needs a complex scalar $\phi$ with SM charges (3,1,-1/3), and three singlet Weyl fermion $N_{i}$. The Lagrangian is given by
\begin{align}
    \label{eq:Lmesino}
    \mathcal{L} \supset y_{ij}\phi \bar{d_{i}} N_{j} - \frac{1}{2} m_{N,ij}N_{i}N_{j} + \alpha_{ij}\phi^{*} \bar{d_{i}}\bar{u_{j}} + c.c.
\end{align}
where $\bar{u_{i}}$ and $\bar{d_{i}}$ are up and down type singlet antiquarks. Neutral mesinos $\Phi_{d_{i}}$ are bound states of $\langle\phi d_{i}\rangle$. Rotating and rephasing the $N_{i}$ makes the singlet mass matrix $m_{N}$ real and diagonal. The first term in Eq.~\ref{eq:Lmesino} is responsible for mesino oscillations mediated via the singlets $N_{i}$. The last term gives the baryon number violating decays of $\phi$ and hence the mesinos. Seven of the nine phases of $\alpha_{ij}$ can be reparametrized and made real. The remaining phases of $\alpha_{ij}$ play no role in contributing to the $CP$ violation required by baryogenesis. This leaves all of the nine complex phases of $y_{ij}$ to give the required $CP$ violation.

Experimental constraints to be discussed below require $m_{\Phi} \simeq m_{\phi} \simeq \mathcal{O}(\TeV)$. We will assume two of the Weyl fermions to be close in mass to the $SU(3)_c-$charged scalar i.e. $m_{N_{i}} = m_{\phi} + \Delta m_{i} $ with $|\Delta m_{i}| \simeq \mathcal{O}(\GeV)$. The Hamiltonian of the mesino - anti-mesino system can be written as
\begin{align}
    \label{eq:Hmesino}
    \mathcal{H} = \begin{pmatrix}
M - i \frac{\Gamma}{2} & M_{12} - i \frac{\Gamma_{12}}{2}\\
M^{*}_{12} - i \frac{\Gamma^{*}_{12}}{2} & M - i \frac{\Gamma}{2}
\end{pmatrix}
\end{align}
where the off diagonal terms $M_{12}, \Gamma_{12}$ are responsible the mesino - anti-mesino oscillations and correspond to on-shell and off-shell contributions respectively. One of the Weyl fermions $N_{1}$ needs to be lighter than the mesino and will contribute to both $M_{12}, \Gamma_{12}$  while $N_{2}$ contributes purely to $M_{12}$. The total baryonic $CP$ asymmetry generated can then be written as
\begin{align}
    \label{eq:mesinoasymmetry}
    \epsilon_{B} = \frac{2 Im(M_{12} \Gamma^{*}_{12})}{\Gamma^{2} + 4 M^{2}_{12}} \frac{\Gamma_{B}}{\Gamma}
\end{align}
where $\Gamma_{B}$ is the width of the mesino decaying to baryons. Generically $\epsilon_{B} \simeq 10^{-3} - 10^{-4}$ but it can be shown that $\epsilon_{B}$ can be as large as $1/8$ since the asymmetry depends on the ratio of values that can be comparable to each other. This is unlike the case where $CP$ asymmetry is generated via decay of heavy particles from interference between their tree level and one-loop diagrams and the $CP$ violating phase has to be large to get a large enough $\epsilon_{B}$.

The cosmology of baryogenesis via mesino oscillations is as follows. The Weyl fermion $N_{3}$ decays into a scalar $\phi$ and a quark at temperatures $T < 200 \MeV$. The scalar $\phi$ quickly hadronizes to form mesinos. $CP$ violation from mesino - anti-mesino oscillations and baryon number violating decays of the mesino result into baryogenesis. The decays of $N_{3}$ and subsequent decay of the mesinos produces entropy which further dilutes the asymmetry. It can be shown that accounting for entropy dilution the maximum possible asymmetry in this model is $\eta_{B,max} \simeq 10^{-6}$, well above the present day baryon asymmetry of $\eta_{B} \approx 10^{-10}$.

Since this model contains a charged $SU(3)_c-$charged scalar it can be produces easily at colliders and this model remains eminently observable. The $SU(3)_c-$charged scalar $\phi$ can decay into two jets through the $\alpha_{ij} \phi^{*} \bar{d_{i}}\bar{u_{j}}$, or appears to decay to two jets via it's decay into $N_{1}$ and the subsequent decay of $N_{1}$ to two hard jets (the third jet is usually soft and not detected at the LHC). However looking for 3 jet events puts a constraint of $m_{\phi} < 600 \GeV$ \cite{ATLAS:2012dp,Aad:2015lea,Chatrchyan:2013gia}. Ref. \cite{Duggan:2013yna} indicates that future 2 jet searches at $14 \TeV$ LHC at $1000^{-1} \rm{fb}$ will constrain $m_{\phi} < \TeV$. Future $100 \TeV$ colliders can probe $m_{\phi} ~ 10 \TeV$. Constraints from searches of displaced vertices at the LHC, neutron - antineutron and kaon oscillations, and from dinucleon decays to kaons are easily satisfied by appropriate choices of $y_{ij}, \alpha_{ij}$. Finally  an observation of decays of mesino, anti-mesino to same sign top quark at the LHC  will be a smoking gun signature for this mechanism \cite{Berger:2012mm}.

\subsection{Mesogenesis}
\subsection*{Contributors: Gilly Elor and Robert McGehee}

\begin{figure}[t!]
    \centering
    \includegraphics[width=\textwidth]{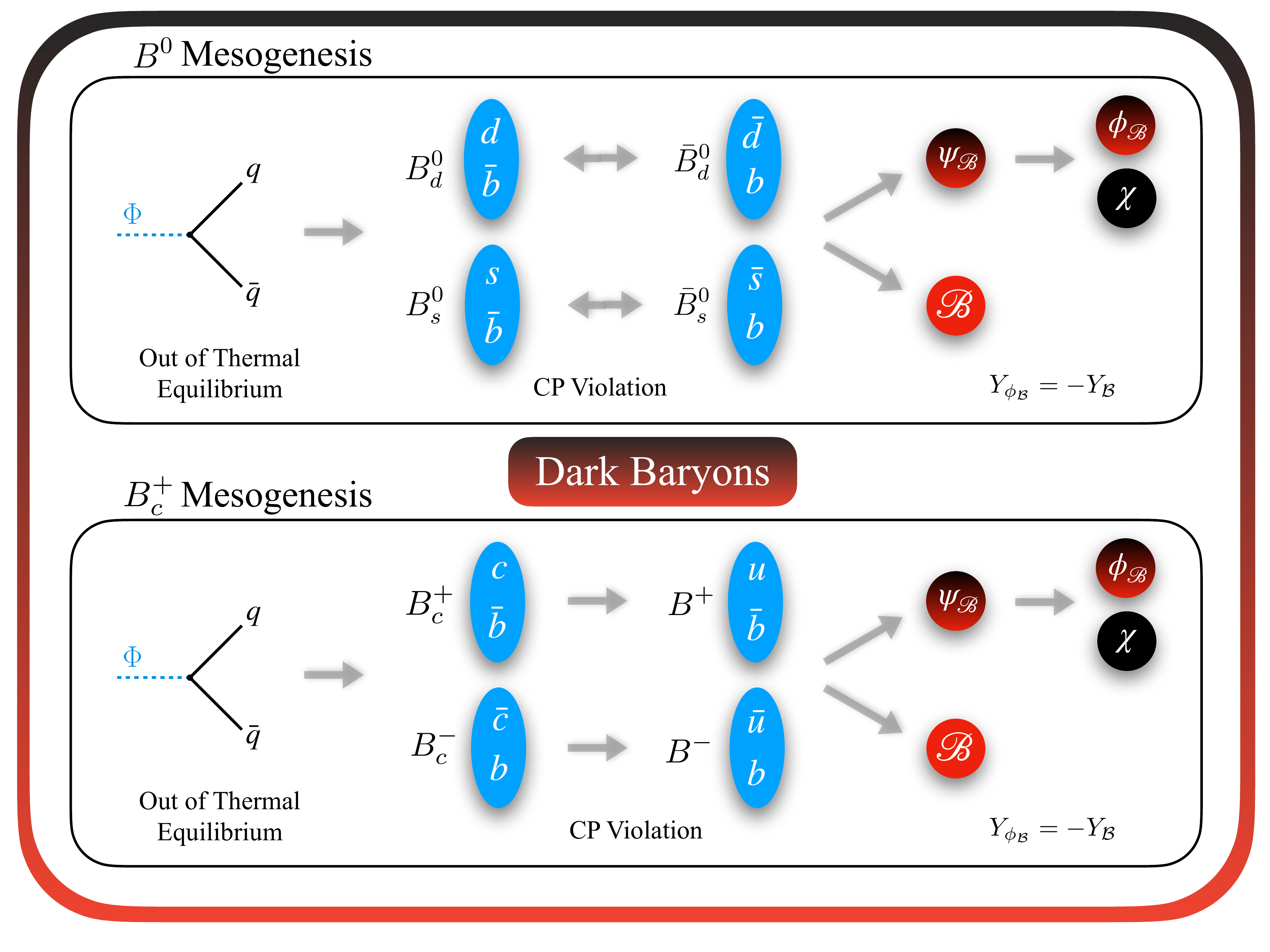}
    \caption{Depiction of how Neutral $B$ Mesogenesis (top) and $B_c^+$ Mesogenesis (bottom) satisfy the Sakhorov conditions generating an equal and opposite baryon asymmetry in the dark and visible sectors.}
    \label{fig:MesoDarkBaryon}
\end{figure}

Mesogenesis is a novel, experimentally testable mechanism of MeV-scale baryogenesis and dark matter production which utilizes $CP$ violation (CPV) within SM meson systems~\cite{Elor:2018twp,Elor:2020tkc,Elahi:2021jia}. 
To date, several mechanisms of Mesogenesis have been proposed. Generic to all \emph{flavors} of Mesogenesis is a scalar field $\Phi$ with a mass of $10-100 \text{ GeV}$ which decays at a low temperature $T_R$ to $q \bar{q}$ pairs. $\Phi$ may or may not be related to inflation, but $T_{\rm BBN} \lesssim T_R \lesssim T_{\rm QCD}$, so there exists a late, matter-dominated era i.e. $\Gamma_\Phi \sim H(T_R)$. At such MeV scales, the $q \bar{q}$'s subsequently hadronize into SM neutral and charged mesons which undergo out-of-equilibrium CPV processes such as neutral $B^0_{d,s}$ oscillations or charged meson decays. These processes are expected in the SM, but CPV contributions from new physics could exist (and is required in some versions of Mesogenesis). Baryon number is never violated thanks to the introduction of a new dark sector state $\psi_{\mathcal{B}}$ carrying baryon number $B=-1$. 

There are two sub-classes of Mesogenesis models. In the first, the daughter meson of the CPV process decays into the dark baryon and a SM baryon generating an equal and opposite baryon asymmetry between the dark and visible sectors. Since the stability of matter requires $m_{\mathcal{B}} \gtrsim m_p$, this is only possible for sufficiently heavy daughter mesons. Fig.~\ref{fig:MesoDarkBaryon} schematically summarizes how the Sakharov conditions are satisfied in such scenarios. In the second sub-class, the daughter meson decays instead into a pair of dark and SM leptons, generating an equal and opposite lepton asymmetry between the dark and visible sectors. This lepton asymmetry is then transferred to a baryon asymmetry between the two sectors via dark-sector processes. This second sub-class, while requiring extra dark-sector dynamics, allows the usage of CPV in lighter SM mesons to generate the baryon asymmetry. These Mesogenesis mechanisms are summarized in Table.~\ref{tab:decayChannels}. In all cases, the generated matter anti-matter asymmetry is directly related to experimental observables such as charge asymmetries or branching fractions of new hadron decay modes. We itemize the relevant observables for each scenario along with the experiments most suited to probe them in Table.~\ref{tab:decayChannels}. Next, we elucidate the details of the various models. 

\begin{table*}[h]
\renewcommand{\arraystretch}{1}
  \setlength{\arrayrulewidth}{.25mm}
\centering
\small
\setlength{\tabcolsep}{0.2 em}
\begin{tabular}{ | c | c | c | c | c  |  c|}
    \hline
  Mechanism & CPV & Dark Sector & Observables  &  Relevant Experiments \\
    \hline \hline
    $B^0$  Mesogenesis 
    &  $B_s^0 \,\, \& \,\, B_d^0$   
    &  dark baryons   
    & $A^{s,d}_{sl}$    
    & LHCb     \\  \cite{Elor:2018twp} 
    &   oscillations
    &  
    & $\text{Br} (B\rightarrow \mathcal{B}+ X)$
    &  $B$ Factories, LHCb    \\
    \hline
    
    &  
    &   
    & $A_{CP}^D$  
    & $B$ Factories, LHCb \\
    $D^+$  Mesogenesis  
    &  $D^\pm$ decays   
    &   dark leptons  
    & $\text{Br}_{D^+}$  
    &  $B$ Factories, LHCb \\ \cite{Elor:2020tkc}
    &    
    &    and baryons
    &  $\text{Br} (\mathcal{M^+} \rightarrow \ell^+ + X)$  
    &  peak searches e.g. PSI, PIENU \\
    \hline
     & 
     & 
     & $A_{CP}^B$
     & $B$ Factories, LHCb\\
    $B^+$  Mesogenesis  
     &  $B^\pm$ decays   
     & dark leptons 
     & $\text{Br}_{B^+}$
     &$B$ Factories, LHCb   \\ \cite{Elahi:2021jia} 
     &   
     &  and baryons
     &  $\text{Br} (\mathcal{M^+} \rightarrow \ell^+ + X)$  
     & peak searches e.g. PSI, PIENU  \\
    \hline
     &  
     &
     & $A_{CP}^{B_c}$ 
     &  LHCb, FCC \\
     $B^+_c$  Mesogenesis 
     &  $B^\pm_c$ decays & dark baryons 
     & $\text{Br}_{B_c^+}$ 
     & LHCb, FCC \\ \cite{Elahi:2021jia}
     &  
     &
     &  $\text{Br}_{B^+\rightarrow \mathcal{B}^++ X}$ 
     &  $B$ Factories, LHCb \\
    \hline

\end{tabular}
\caption{Summary of different flavors of Mesogenesis. Indirect signals are not shown, but are discussed in the text. 
}
\label{tab:decayChannels}
\end{table*}

\begin{figure}[t!]
    \centering
    \includegraphics[width=\textwidth]{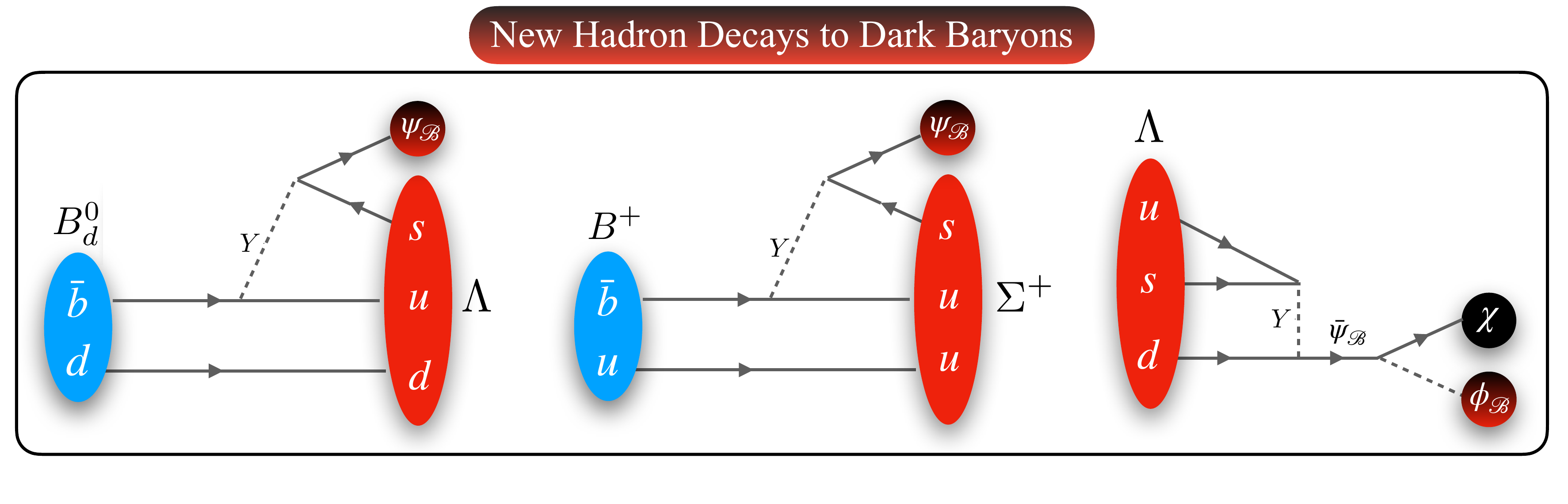}
    \caption{A few examples of new decay modes of SM hadrons to dark baryons. In the left and middle diagrams, the $B$ meson decay proceeds through $\mathcal{O}_{bus}$. The left decay $B_d^0 \rightarrow \Lambda^0 \psi_{\mathcal{B}}$ is directly related to the baryon asymmetry in neutral $B$ Mesogenesis, while the middle diagram $B^+ \rightarrow \Sigma^+ \psi_{\mathcal{B}}$ is an observable directly related to $B_c^+$ Mesogenesis. In the first, a designated search targeting Mesogenesis by the Belle Collaboration \cite{Belle:2021gmc} has probed this operator (see text for implications). In the right-most diagram, the fully invisible decay of the $\Lambda$ baryon proceeds through the $\mathcal{O}_{usd}$ operator and indirectly probes neutral $B$ and $B_c^+$ Mesogenesis. A recent search by BESIII \cite{BESIII:2021slv} has targeted exactly this decay mode. See \cite{Alonso-Alvarez:2021oaj} for other $s$-flavored baryon decays arising in this model.}
    \label{fig:DarkBaryonDecays}
\end{figure}

\subsubsection*{Neutral $B$ Mesogenesis}
In neutral $B$ Mesogenesis \cite{Elor:2018twp}, the CPV of $B_{s,d}^0-\bar{B}_{s,d}^0$ oscillations is leveraged. To mediate the decay into the dark state, one introduces a $SU(3)_c-$charged triplet scalar $Y$ with electric charge $-1/3$ and baryon number $-2/3$. The following interactions are allowed 
\bea
\mathcal{L}_Y = - \sum_{i,j} y_{ij} Y^* \bar{u}_{i,R} d_{j,R}^c - \sum_k y_{\psi_{\mathcal{B}}k} Y \bar{\psi}_{\mathcal{B}} d_{kR}^c  + \text{h.c.}
\label{eq:UVmodel}
\eea
Consistency with LHC bounds requires $M_Y \sim \mathcal{O}(\rm TeV)$, so integrating out this scalar yields the effective operator: 
\bea
\label{eq:EffOp}
\mathcal{O}_{b u_i d_j} = \frac{y^2}{M_Y^2} \bar{u}_i^c d_j \bar{b}^c \psi_{\mathcal{B}}    + \text{h.c.}
\eea
where $y^2 \equiv y_{ij} y_{\psi_{\mathcal{B}} 3}$. This allows the $b$ quark within the meson to decay 
via $\bar{b} \rightarrow \psi_{\mathcal{B}} u d$. After undergoing oscillations, the $B$ meson decays into dark and SM baryons, resulting in an equal and opposite baryon asymmetry between the dark and visible sectors. The baryon asymmetry is directly linked to experimental observables and successful Mesogenesis requires \bea
\label{eq:neutralBparamspace}
A_{\rm sl} \times \text{Br}\left( B^0 \rightarrow \mathcal{B}_{\rm SM} + \bar{\psi}_{\mathcal{B}} \right) \gtrsim 10^{-7} \,,
\eea
where $A_{\rm sl}$ is the semi-leptonic asymmetry.  Searches for the apparent baryon-number-violating meson decays in this mechanism are already underway at Belle~\cite{Belle:2021gmc} and LHCb \cite{Rodriguez:2021urv,Borsato:2021aum}.
In particular, in ~\cite{Belle:2021gmc} a designated search using Belle-I data was performed that set a bound $\text{Br} \left( B^0 \rightarrow \Lambda^0 + \psi_{\mathcal{B}} \right) \lesssim 2.1-3.8 \times 10^{-5}$. Given the requirement for successful neutral $B$ Mesogenesis Eq.~\eqref{eq:neutralBparamspace} it is clear that this branching fraction is too small to generate the requisite baryon asymmetry. However, the three other decay modes corresponding to flavorful variations of Eq.~\eqref{eq:EffOp} (see e.g. Table I of \cite{Alonso-Alvarez:2021qfd}) are still viable and measuring these branching fractions would be a direct probe of neutral $B$ Mesogenesis. 

The UV model which gives rise to Eq.~\eqref{eq:EffOp} also predicts new decay modes for strange baryons \cite{Alonso-Alvarez:2021oaj} which can be probed at Kaon and Hyperon factories \cite{Goudzovski:2022vbt}. BESIII has recently set limits on the fully invisible decay of $\Lambda$ baryons \cite{BESIII:2021slv}. Since neutral $B$ Mesogenesis predicts the existence of a $SU(3)_c-$charged triplet mediator, collider and flavor observables can indirectly probe this mechanism. Fig.~\ref{fig:DarkBaryonDecays} illustrates a few of the new hadron decays that arise from Eq.~\ref{eq:EffOp} which are either directly or indirectly related to the baryon asymmetry in this framework. Given the plethora of signals and ongoing experimental searches, neutral $B$ Mesogenesis is likely to be fully probed within the next 5-10 years (see Fig. 3 of \cite{Alonso-Alvarez:2021qfd} for the projected limits on the viable parameter space).

Since the dark baryon $\psi_{\mathcal{B}}$ is required to be sufficiently heavy to kinematically forbid proton decay, the operator Eq.~\eqref{eq:EffOp} would allow any produced $\psi_{\mathcal{B}}$ to decay back into light SM baryons thereby washing out the generated asymmetry. To evade this we must minimally extend the dark sector such that the $\psi_{\mathcal{B}}$'s instead quickly decays into the dark sector. We introduce two new dark sector states; a scalar baryon $\psi_{\mathcal{B}}$, and a dark fermion $\chi$ both odd under a $\mathbb{Z}_2$ symmetry. The following Lagrangian is allowed $\mathcal{L} \supset y_{\rm dark} \, \psi_{\mathcal{B}} \phi_{\mathcal{B}} \chi$, and mediates the decay $\psi_{\mathcal{B}} \rightarrow \phi_{\mathcal{B}} + \chi$.  The produced $\phi_{\mathcal{B}}$'s and $\chi$'s can make up the dark matter with relic abundance $\Omega h^2 = 0.11$.

Neutral $B$ Mesogenesis can also be explicitly realized in a supersymmetric model with Dirac Gauginos and an $R$-symmetry identified with baryon number \cite{Alonso-Alvarez:2019fym}. The dark matter is associated with a sterile sneutrino multiplet that carries both lepton number and $R$ charge. In this set-up, additional model dependent signals in \emph{e.g.} neutrino experiments arise.

\subsubsection*{$B_c^+$ Mesogenesis}
In $B_c^+$ Mesogenesis \cite{Elahi:2021jia}, $B_c^+$ undergoes a CPV decay to $B^+$ which subsequently decays into the dark sector via the operator in Eq.~\eqref{eq:EffOp}: 
\begin{align}
\label{eq:BcMech}
B_c^+ \to \, & B^+ + f \,, \, \quad  B^+ \to \, \psi_{\mathcal{B}} + \mathcal{B}^+.
\end{align}
The baryon asymmetry is directly controlled by 1) the CPV in $B_c^+$ decays, 2) the branching fraction of the $B_c^+$ decay into $B^+$ mesons and other SM final states, and 3) the branching fraction of the $B^+$ meson into SM baryons and missing energy. The first observable is expected to be sizeable~\cite{Choi:2009ym} but is currently not well constrained, nor is the second. However, the branching fraction of $B^+$ is being probed by the same searches as neutral $B$ Mesogenesis. In particular, the results from the search performed by Belle \cite{Belle:2021gmc} can be recast to set a limit $\text{Br} \left(B^+ \rightarrow \Sigma^+ \bar{\psi}_{\mathcal{B}} \right) \lesssim 2\times 10^{-5}$, which we note is still a viable channel to generate the baryon asymmetry through $B_c^+$ Mesogenesis (see Fig. 2 of \cite{Elahi:2021jia}). Overall, this is a remarkably simple model of Mesogenesis and provides motivations for $B_c$ physics searches at e.g. the LHCb~\cite{Gouz:2002kk} and an electron Future Circular Collider~\cite{FCC:2018byv}.

\subsubsection*{$B^+$ and $D^+$ Mesogenesis}
\begin{figure}[t!]
    \centering
    \includegraphics[width=\textwidth]{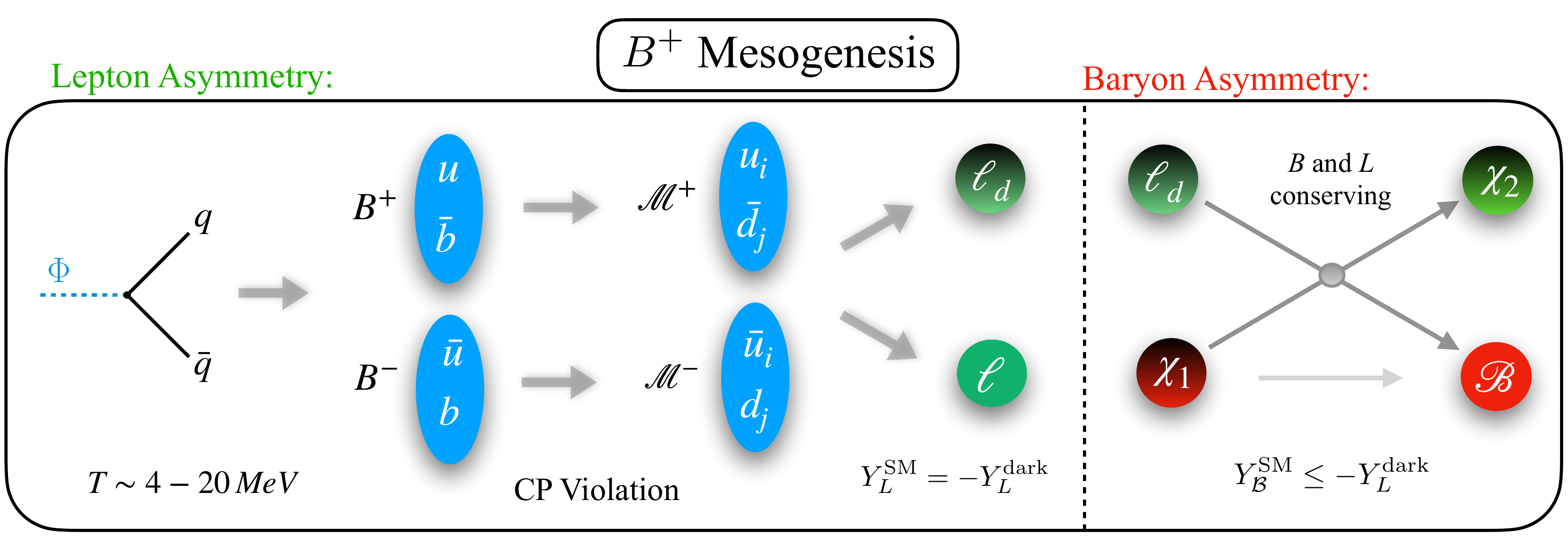}
    \caption{Depiction of how $B^+$ Mesogenesis generates a lepton asymmetry at low scales which is then transferred to a baryon asymmetry. $D^+$ Mesogenesis proceeds in the same way with $B \rightarrow D$.}
    \label{fig:MesoDarkLepton}
\end{figure}

In the second sub-class of Mesogenesis models, the daughter mesons of the CPV process are too light to decay to a pair of dark and SM baryons. Instead, they decay into a pair of dark and SM leptons resulting in an equal and opposite lepton asymmetry between the dark and visible sectors.  A depiction of this mechanism is shown in  Fig.~\ref{fig:MesoDarkLepton}. Two such models of Mesogenesis involve CPV decays of $D^+$~\cite{Elor:2020tkc} and $B^+$~\cite{Elahi:2021jia} mesons:
\begin{align}
 D^+ \, \text{or} \, B^+ \, \to \, &\mathcal{M}^+ \,+\,  \mathcal{M} \,, \quad
\mathcal{M}^+  \, \to \, \ell_d \,+\, \ell^+ \,, 
\end{align}
where $\mathcal{M}^+$ is a charged SM meson. Since this process occurs at MeV scales, electroweak sphalerons cannot convert this lepton asymmetry into a baryon asymmetry, but dark-sector scattering can sufficiently transfer the lepton asymmetry to a baryon asymmetry. As above, the generated lepton asymmetry is directly tied to experimental observables such as the CPV in a particular decay mode:
\begin{align}
A_{\rm CP} = \frac{\Gamma(D^+ \rightarrow f) - \Gamma(D^- \rightarrow \bar{f})}{\Gamma(D^+ \rightarrow f) + \Gamma(D^- \rightarrow \bar{f})}
\end{align}
(and the analogous definition for $B^+$ decays). To achieve a lepton asymmetry greater than the observed baryon asymmetry, the relevant CPV and branching ratios in each Mesogenesis model must satisfy
\begin{subequations}
\begin{align}
D^+: \quad \sum_{f \supset \pi^+} A_{\rm CP}^f \text{Br}(D^+ \to f) \gtrsim 3\times 10^{-5}, \quad \text{Br}(\pi^+ \to \ell_d + \ell^+) \gtrsim 10^{-3},\\
B^+: \quad \sum_{f \supset \mathcal{M}^+} A_{\rm CP}^f \text{Br} (B^+ \to f) \gtrsim 5.4 \times 10^{-5}, \quad \sum_{\mathcal{M}^+} \text{Br} (\mathcal{M}^+ \to \ell_d + \ell^+) \gtrsim 10^{-3}.
\end{align}
\end{subequations}
$D^+$ Mesogenesis may thus be probed by improved sensitivity to both CPV and branching ratios of $D^+$ decays to pions (at \emph{e.g.} LHCb) and $\text{Br}(\pi^+ \to \ell_d + \ell^+)$. 

In $B^+$ Mesogenesis, it may be possible that the SM contains the necessary CPV and branching ratios required to produce the observed baryon asymmetry. Although it is difficult to calculate $A_{\rm CP}^f$, some predicted branching fractions of $B^+$ are on the order of the current experimental central values~\cite{Beneke:2005vv}. It is instead easier to probe the decays of the lighter $\mathcal{M}^+$ to SM leptons $+$ invisible (\emph{e.g.}~\cite{Hayano:1982wu,E949:2014gsn,Aguilar-Arevalo:2017vlf,NA62:2017qcd,Aguilar-Arevalo:2019owf,NA62:2021bji}), often by recasting searches for sterile neutrinos.

\subsection{Particle Asymmetries from Quantum Statistics}
\subsection*{Contributer: Nikita Blinov}
Most common baryogenesis and leptogenesis mechanisms implement $CP$ violation in a way that is directly tied to the other Sakharov conditions. For example, in EW baryogenesis $CP$ violation occurs at the expanding bubble walls that generate an out-of-equilibrium situation. In Ref.~\cite{Blinov:2017dap} we explored an alternative possibility where $CP$ violation is completely sequestered from the other aspects of baryogenesis. We studied the scenario where a particle asymmetry is produced in a dark sector by an unspecified mechanism, leading to a population of an asymmetric, stable species, i.e. asymmetric dark matter (ADM). Unlike standard ADM, however, the dark matter asymmetry is not shared with the visible sector directly via transfer operators, but rather via quantum statistical effects. In other words, the presence of an asymmetric species biases an out-of-equilibrium, visible baryon-number violating process to produce more baryons than anti-baryons.

As a simple example of this mechanism, consider the out-of-equilibrium decays of a real scalar $\varphi$ with the interaction
\bea
\mathcal{L} \supset \frac{1}{\Lambda} \varphi \psi_B \psi_D \phi^\dagger_D
+ \hc
\label{eq:bose_model}
\eea
where $\psi_B$ is a fermion that carries baryon number and $\psi_D$
($\phi_D$) is a fermion (scalar) that carries a $U(1)_D$ dark quantum
number ($\psi_B$ would need to be converted to SM baryons via, e.g., the neutron portal $\psi_B u^c d^c d^c$). This interaction gives rise to two decay channels for $\varphi$, $\varphi \to \psi_B \psi_D \phi^\dagger_D$ and $\varphi \to \psi_B^\dagger \psi_D^\dagger \phi_D$.
The decays of $\varphi$ violate baryon
number but preserve dark matter number. In the absence of any $CP$ violation,
Lagrangian or otherwise, these two decays have equal probabilities so that no
baryon number asymmetry is generated.

At finite temperature the
rate for a single decay channel is 
\beq
\Gamma(\varphi\rightarrow \psi_B \psi_D \varphi_D^\dagger) = 
\frac{1}{2M_\varphi} \int d\Phi_3 |\mathcal{M}|^2 
(1 + f_{\phi_D})
(1 - f_{\psi_D})
(1 - f_{\psi_B}),
\label{eq:basic_decay_rate}
\eeq
where $\mathcal{M}$ is the decay matrix element, $d\Phi_3$ is the three-body
phase space volume element, and $f_i$ are phase-space distributions of the different species. From this definition it is clear that the decay asymmetry
\beq
\Delta\Gamma =
\Gamma(\varphi\rightarrow \psi_B \psi_D \phi_D^\dagger) - \Gamma(\varphi\rightarrow \bar \psi_B \bar \psi_D \phi_D)
\label{eq:decay_asym_def}
\eeq
will be non-zero if $f_{\phi_D}\neq f_{\phi_D^\dagger}$ or $f_{\psi_D}\neq f_{\bar\psi_D}$. In fact, one can show analytically that the decay asymmetry is proportional to the DM chemical potential, and it is temperature-dependent. This leads to the majority of the asymmetry being produced before $\varphi$ decays away completely, in contrast to standard out-of-equilibrium decay scenarios with a constant $\Delta \Gamma$. Viable baryogenesis can be achieved with a dark matter asymmetry similar to, or much larger than the visible asymmetry.
This means that if the ADM is the dark matter today, its mass relation to proton mass can be radically different compared to the standard ADM expectation of $m_{DM} \sim m_p$. 

The sequestration of $CP$ violation from baryogenesis means that this type of model can be directly tested only through the operators connecting the visible baryons to new state $\psi_B$. 

\section{New Ideas for Testing Traditional Mechanisms}
\label{sec:NewTests}
In the decades following Sakharov's paper, a variety of mechanisms for baryogenesis were proposed. However, many of these proposals, such as Leptogenesis and Affleck-Dine baryogenesis, involved high scales and very massive particles, making them notoriously difficult to test experimentally. In recent years, novel proposals for experimentally testing more traditional models have been put forth. In this section we highlight several of these proposals.

\subsection{Prospects for Detection of Affleck-Dine Baryogenesis}
\subsection*{Contributors: Fatemeh Elahi and Graham White}

In the supersymmetric extension of the SM, the scalar potential generally has several flat directions, allowing scalar condensate to develop large vacuum expectation values (vevs) during inflation. Soft supersymmetry breaking terms and higher order non-renormalizable terms, however, lift the flat directions and induce an oscillation around the minimum during post-inflationary epoch~\cite{Gherghetta:1995dv}. Affleck and Dine proposed a novel scenario for baryogenesis utilizing these flat directions, focusing on the case where the scalar potential carries a non-zero baryon or lepton number~\cite{Affleck:1984fy}. The non-zero vev of the condensate determines the initial position of the field and the baryon number violating terms in the potential result in the rotating trajectory for the vev for the evolution of the condensate. If the trajectory has a non-zero baryon number, then the condensate carries a baryon asymmetry, which will transfer to baryons during its decay. As the temperature decreases, the effects of B-violating terms become negligible and the amount of asymmetry stays constant. The details of this scenario has been investigated extensively ($e.g.$, 
~\cite{Affleck:1983mk,Affleck:1984uz,Dine:1995uk,Dine:1995kz,Kolda:1998kc,Anisimov:2000wx,Enqvist:2003gh,Cheung:2012im}. Also, see~\cite{Riotto:1999yt,Allahverdi:2012ju} and references therein for a more extensive review). A similar setup would also work if the phase transition was first order \cite{Cheung:2012im} resulting in asymmetry generated in the bubble walls and subsequent collisions.

Given that the weak sphalerons can wash-out the baryon asymmetry very rapidly before the electroweak phase transition (EWPT) ~\cite{Arnold:1996dy,DOnofrio:2012yxq}, extra complications are needed to preserve the asymmetry till EWPT. One idea is making sure the condensate decays very late. However, in this case, the coherent oscillation of the field needs to last long, and thus its potential needs to be immune from various fluctuations. A prevailing example of such a scenario is Q-ball (e.g, 
~\cite{kusenko1997solitons,ENQVIST1998309,Kasuya:2000wx,Fujii:2002kr,allahverdi2002thermalization,
Shoemaker:2009kg}). Another variation, which tames the work of the weak sphalerons, is assuming a non-zero seed of an anomalous hypermagnetic field (AHMF). It has been shown that an AHMF with a large amplitude can save the asymmetry from being washed-out, because of the chiral magnetic effect ~\cite{Semikoz:2011,Dvornikov:2013,Kuzmin:1985mm,Zadeh:2018usa,Long:2014,Rubakov:1986am,Giovannini:1998,Giovannini:1998b,Joyce:1997,Khlebnikov:1988sr,Zadeh:2016nfk,Zadeh:2015oqf,Mottola:1990bz,Abbaslu:2019yiy,Elahi:2020pxl,Elahi:2021pug}.

One promising way to test Affleck Dine baryogenesis is through secondary gravitational wave production via the poltergeist mechanism \cite{White:2021hwi}. The $B$ and $L$-violating field space directions directions involved in Affleck dine baryogenesis are flat and therefore typically allow for the production of Q-balls, with either $B$ or $L$ playing the role of the conserved global charge. These flat directions are not expected to couple strongly to any other field, as loop corrections to the potential would then jeopardize the flatness of the potential. This means the particle decay of the Affleck Dine condensate is expected to be slow and the system usually prefers to fragment into Q-balls. Simulations suggest a large symmetric component to the energy \cite{Hiramatsu:2010dx} which means that generally the amount of energy in the Q-balls isn't too many orders of magnitude smaller than the radiation component of the energy density. The Q-balls redshift like matter, so if they survive long enough they will come to dominate the energy. Typically, the Q-balls cannot decay into fermions, except at the surface, as even if such a decay is kinematically allowed, the Fermi sea quickly fills up and Pauli blocking prevents decay. This fact conspires to make the Q-balls typically long lived enough to dominate the energy density. When the Q-balls begin to decay, the process accelerates such that the decay is faster than a Hubble time \cite{Cohen:1986ct}. The Universe will therefore undergo a sudden change in the equation of state which leads to a resonant enhancement in gravitational waves \cite{Inomata:2019ivs}, usually at a low enough frequency to detect. Therefore, Affleck Dine baryogenesis typically results in a gravitational wave signal that is usually detectable and can only be produced by a limited number of cosmological scenarios.

\subsection{Probing High-Scale Baryogenesis with Neutron-Antineutron Oscillations}

\subsection*{Contributors: K\aa{}re Fridell, Julia Harz, Chandan Hati and Bibhushan Shakya}

Searches for baryon and lepton number ($B$ and $L$) violation are not only powerful probes of physics beyond the SM but also can provide valuable insight into the mechanism responsible for generation of the observed baryon asymmetry of the Universe. If baryogenesis occurs at a temperature above the electroweak breaking scale, then an asymmetry must be produced via $B-L$ violation to avoid washout from electroweak sphalerons, which conserve $B-L$ but violate $B+L$. An interesting possibility to consider is direct $B$ violation, which can be probed with observables violating $B$ and $B-L$, but conserving $L$. Neutron-antineutron ($n-\bar{n}$) oscillations provide such an observable, which can be described by dimension-9 effective operators with $|\Delta B|=2$ and $\Delta L=0$. Current measurements constrain the $n-\bar{n}$ oscillation lifetime to $\tau_{\text{free}}\geq 0.86 \times 10^8$~s for free $n-\bar{n}$ oscillation~\cite{Baldo-Ceolin:1994hzw} and $\tau_{\text{bound}}\geq 4.7 \times 10^8$~s for bound $n-\bar{n}$ oscillation~\cite{Super-Kamiokande:2020bov}. Interestingly, the upcoming NNBAR experiment at the ESS facility and DUNE will improve on these limits by orders of magnitude~\cite{Addazi:2020nlz,DUNE:2015lol} and therefore can provide a first smoking gun signal for direct $B$ violation. Such a signal can have far-reaching consequences and features interesting complementarities with other observables at the high-energy and high-intensity frontiers. In particular, we can distinguish two cases:

(1) \textbf{$n-\bar{n}$ oscillations featuring new physics without CPV interaction:} From a model independent point of view, if the new physics fields mediating $n-\bar{n}$ oscillations are significantly heavier than the external quarks in the effective operators corresponding to $n-\bar{n}$ oscillation, then at any temperature below the mass scale of new physics the effective $n-\bar{n}$ oscillation operators correspond to $|\Delta B|=2$ wash-out processes removing any pre-existing $B-L$ asymmetry. Therefore, an observed rate of $n-\bar{n}$ oscillations at low-energy experiments directly implies a certain washout rate of a pre-existing asymmetry. A simple estimate shows that an observation of $n-\bar{n}$ oscillations at upcoming experiments like DUNE or NNBAR would imply that washout processes corresponding to $n-\bar{n}$ oscillation operators will remain in equilibrium till around 100 TeV~\cite{Fridell:2021gag}. Hence, in order to generate the observed baryon asymmetry, baryogenesis must occur below the 100 TeV scale, making the relevant new physics fields detectable at a future 100 TeV collider. Alternatively, it might point to a baryogenesis mechanism restricted to second or third generations of quarks only or a secluded flavour sector.

(2) \textbf{$n-\bar{n}$ oscillations featuring new physics with CPV interaction:} At tree level, there are two topologies that can UV-complete the dimension-9 $n\--\bar{n}$ effective operators, one with a trilinear boson coupling~\cite{Mohapatra:1980qe,Mohapatra:1982xz,Chang:1984qr,Babu:2006xc,Babu:2008rq,Baldes:2011mh,Babu:2012vc,Arnold:2012sd,Babu:2013yca,Patra:2014goa,Herrmann:2014fha} and one with two bosons and a Majorana fermion~\cite{Zwirner:1984is,Barbieri:1985ty,Mohapatra:1986bd,Lazarides:1986jt,Goity:1994dq,Babu:2001qr,Babu:2006wz,Allahverdi:2010im,Gu:2011ff,Gu:2011fp,Allahverdi:2013mza,Dev:2015uca,Dhuria:2015swa,Ghalsasi:2015mxa,Gu:2016ghu,Calibbi:2016ukt,Gu:2017cgp,Calibbi:2017rab,Allahverdi:2017edd}. Depending on the relevant heavy new physics, a $CP$-violating out-of-equilibrium decay of new heavy fields can lead to a baryon asymmetry generation, while inducing different signals at collider experiments, in dinucleon decay, and meson oscillations. 

A detailed discussion of the topology with two bosons and a Majorana fermion in the context of baryogenesis can be found in~\cite{Grojean:2018fus}. The production of the baryon asymmetry requires two Majorana fermions, with broadly two viable classes of scenarios: (i) relatively early decay of the heavier fermion, along with suppressed washout effects due to weaker couplings of the lighter fermion; the heavier fermion dominates the $n-\bar{n}$ oscillation signal, and projected sensitivities can probe masses in the $10-1000$ TeV regime; (ii) late decay of the heavier fermion after washout effects have gone out of equilibrium; in this case the lighter fermion dominates the $n-\bar{n}$ oscillation signal, and projected sensitivities can probe masses in the $1-1000$ TeV regime. Such topologies can be realized in supersymmetric models (see e.g. \cite{Cui:2012jh,Cui:2013bta,Arcadi:2015ffa,Pierce:2019ozl}), where the two bosons correspond to squarks, and the Majorana fermion corresponds to a gaugino, with the necessary couplings arising from R-parity violating operators. 

In the case where the trilinear topology and the baryogenesis mechanism is realised via two heavy new bosons with similar masses e.g. by two scalar diquarks, the model independent effective operator results are recovered: A signal of $n-\bar{n}$ oscillations (corresponding to $m_{\text{DQ}}< \text{few}\, 100~\mathrm{TeV}$) at the upcoming experiments would rule out the possibility of baryogenesis due to very strong washout effects even if the $CP$ violation considered is maximal~\cite{Fridell:2021gag}. In contrast, if the new diquark states are largely hierarchical in their masses (we call this the high-scale scenario, a scenario which is beyond the validity of the simple effective field theory approach mentioned before) with one of the diquarks at near grand unification scale and the other within future collider reach, then an interesting interplay between the LHC and $n-\bar{n}$ oscillation searches can constrain the relevant parameters of this baryogenesis scenario. If a diquark signal is observed at the LHC through third or lighter generation quark final states, then $n-\bar{n}$ oscillation experiments can give complementary insight into the flavour dynamics of high-scale baryogenesis scenario, and would set the stage for exploring flavour effects. If a signal at both the LHC and $n-\bar{n}$ oscillation experiments is discovered, then naively considering the maximal washout corresponding to the first generation $n-\bar{n}$ oscillation rate in a single flavoured analysis suggests that the high-scale scenario remains a viable and attractive option for baryogenesis~\cite{Fridell:2021gag}. 

\subsection{Probing High-Scale Leptogenesis with TeV-scale Lepton-Number Violation}
\subsection*{Contributor: Julia Harz}

Besides generating the observed baryon asymmetry directly via $B$-violating decays, another established possibility is to first generate a lepton asymmetry that then gets directly transferred into a baryon asymmetry via the SM sphaleron processes. On the one hand, the standard leptogenesis mechanism, namely the out-of-equilibrium decay of heavy right-handed neutrinos (RHNs) featuring $CP$-violating interactions, provides a possible link to the mechanism behind the neutrino masses. On the other hand, it is difficult to probe as the typically expected RHN masses~\cite{Davidson:2002qv} exceed the energy reach of current and future colliders (except e.g. resonant leptogenesis~\cite{Pilaftsis:2003gt}). 

Therefore, it is useful to explore with which approaches one is able to exclude certain scenarios or models. The observation of TeV-scale lepton-number violating interactions either from meson decays, neutrinoless double beta decay or same-sign dilepton signatures at the LHC or future colliders has far-reaching consequences in this context.

As was shown in~\cite{Deppisch:2013jxa}, the observation of a lepton-number violating dilepton signature without missing energy at the LHC would directly imply such a strong washout that an asymmetry generated at a high scale would have been completely depleted, rendering high-scale leptogenesis unviable. For mid-scale scenarios (e.g. resonant leptogenesis) such a measurement would imply a lower bound on the $CP$ asymmetry needed to generate the observed baryon asymmetry.

Similarly, an observation of neutrinoless double beta decay based on new physics that can be described by an effective operator of dim-7 (long-range contribution), dim-9 (short-range contribution) or higher would imply an exclusion of high-scale leptogenesis scenarios~\cite{Deppisch:2015yqa,Deppisch:2017ecm}. Therefore, after a discovery the identification of the underlying new physics and hence its effective description should have high priority~\cite{Deppisch:2006hb,Gehman:2007qg}. However, an observation of LNV in all flavours (e.g. at colliders) or an additional measurement of lepton-flavour violation is required in order to confirm washout not only in the first generation.

Therefore, the complementarity of collider searches and neutrinoless double beta decay experiments is particularly important in this endeavour. As shown in~\cite{Harz:2021psp}, depending on the hierarchy of the new physics involved, either collider experiments or neutrinoless double beta decay experiments have the larger reach with respect to first generation leptons. However, in contrast to neutrinoless double beta decay experiments, collider experiments also allow for searches of same-sign dileptons of second and third generation. Generally, observing a TeV-scale lepton-number violating signal in one of those experiments would render single-flavour standard leptogenesis invalid. For a final conclusion equilibration in all flavours has to be confirmed.

Important to note is that due to the $B-L$ conserving sphaleron interactions, a lepton-number violating signature might put high-scale baryogenesis models similarly under tension, as due to the electroweak sphaleron processes a strong lepton washout also implies a strong baryon washout. However, caveats might apply for certain types of models. For instance, models that hide an asymmetry in a specific flavour or that generate the asymmetry in the context of a dark symmetry~\cite{AristizabalSierra:2013lyx,Frandsen:2018jfi} might be exempt of the washout that arises from the observable processes. This demonstrates that experimental searches for lepton-number violating observables are of high relevance and should thus be undertaken with high priority.


\subsection{Cosmological Collider Signals of Leptogenesis}
\subsection*{Contributor: Yanou Cui}

Despite its theoretical appeal, the leptogenesis mechanism is rather challenging to directly test due to the very high energy scales involved. In a recent work \cite{Cui:2021iie} a novel probe for leptogenesis with cosmological collider (CC) physics was proposed. Cosmological collider physics has been developed in recent years as a new method for probing new heavy particles taking advantage of the huge energy available during cosmic inflation, which can be up to $O(10^{13})~{\rm GeV}$~ \cite{Baumann:2011nk,Arkani-Hamed:2015bza,Chen:2016nrs,Lee:2016vti,Chen:2016uwp,Chen:2016hrz,An:2017hlx,Kumar:2017ecc,Chen:2017ryl,Chen:2018xck,Wu:2018lmx,Li:2019ves,Lu:2019tjj,Liu:2019fag,Hook:2019zxa,Hook:2019vcn,Kumar:2019ebj,Alexander:2019vtb,Wang:2019gbi,Wang:2019gok,Wang:2020uic,Li:2020xwr,Wang:2020ioa,Fan:2020xgh,Aoki:2020zbj,Bodas:2020yho,Maru:2021ezc,Lu:2021wxu,Wang:2021qez,Kim:2021ida}. \cite{Cui:2021iie} demonstrated a new application of this approach in testing high scale baryogenesis models such as leptogenesis. Given that SM Higgs directly participates in leptogenesis via the Yukawa coupling, \cite{Cui:2021iie} focuses on the scenario of cosmological higgs collider \cite{Lu:2019tjj,Dvali:2003em, Kofman:2003nx, Suyama:2007bg, Ichikawa:2008ne}, where SM Higgs contributes to generating primordial fluctuation. By computing the bispectrum (i.e. the 3-point correlation function) of the Higgs field fluctuation during inflation in the squeezed limit, it was found that for viable leptogenesis models the amplitude of the primordial non-Gaussianity $f_{NL}$ can be within reach of the upcoming CMB/LSS/21 cm line experiments. Furthermore, the oscillation pattern of the shape function of Higgs bispectrum distinguishes itself from existing results in the CC literature. This results from the necessary $CP$ phase in the Yukawa couplings as well as the large quantum fluctuation of the Higgs field during inflation, which leads to Yukawa couplings mixing different neutrino mass eigenstates with comparable masses. To summarize, this work presents an intriguing case that dedicated measurement of primordial non-Gaussianity could shed light on high scale leptogenesis by revealing key information about the lepton-number violating couplings, the Majorana right-hand neutrino masses, and the $CP$ phases.

 \subsection{Cosmic Strings and Tests of Thermal Leptogenesis}
 \subsection*{Contributor: Graham White}
 
 There is expected to be a hierarchy between the scale of $B-L$ breaking from the sterile mass operator and the scale of grand unification. Therefore it is natural to consider a symmetry to protect the $B-L$ breaking scale, that is we can gauge $B-L$ and embed this symmetry in the larger GUT. One can consider all symmetry breaking paths involving groups that are at most rank 5 and non-anomolous with SM fermions and sterile neutrinos only. Inflation must occur after monopoles are produced, to be consistent with their non observation, but before $B-L$ breaking where thermal leptogenesis occurs. The majority of symmetry breaking paths that are viable (that is, do not have stable domain walls) predict cosmic strings. The gravitational wave signal for such strings has a large enough amplitude that future detectors are expected to probe the entire range relevant for thermal leptogenesis \cite{Dror:2019syi}.
 
 \subsection{Vacuum Instability Tests of the Minimal Leptogenesis Scenario}
  \subsection*{Contributor: Graham White}
 A nightmare scenario for baryogenesis is that it is explained by a bare minimal extension of at least two sterile neutrinos. In this case testability becomes very difficult. The one probe that could in principle still be useful is tests of vacuum stability. There are two ways where the vacuum stability affects the parameter space of thermal leptogenesis. First, even if the stability scale is below the scale of sterile neutrinos, the negative contribution to the beta function of the Higgs self coupling causes more efficient tunneling \cite{Ipek:2018sai}. This means that the trace of the Yukawa couplings, and therefore the mass of the heavier steriles have a non trivial upper bound. Second, during reheating any couplings between the inflaton and the Higgs can cause resonant production resulting in catastrophic vacuum decay \cite{Enqvist:2016mqj}. In a a minimal model such couplings are inevitably produced radiatively. This results in a strong constraint on the leptogenesis parameter space in the plane of the sterile mass and the reheating temperature \cite{Croon:2019dfw}.
 
\subsection{First Order Phase Transitions}
  \subsection*{Contributor: Peisi Huang, Jorinde van de Vis}
In the SM, electroweak symmetry breaking~(EWSB) would proceed via a smooth crossover unless the Higgs mass is below $\sim$70 GeV~\cite{Dine:1992vs,Kajantie:1995kf}. Therefore, the discovery of the SM Higgs boson with a mass $m_h = 125$~GeV~\cite{ATLAS:2012dp,CMS:2012qbp} meant that the SM alone cannot satisfy the third Sakharov condition, i.e., departure from thermal equilibrium. This has motivated a further investigation of the viability of Electroweak baryogenesis in minimally extended scenarios.

 A modification of the nature of the electroweak phase transition~(EWPT) may be achieved by adding new interactions to the Higgs potential~\cite{Cohen:1993nk,Riotto:1999yt,Morrissey:2012db,Huang:2015tdv}. These may result in relevant temperature dependent modifications to the Higgs potential, beyond those associated with the increase of the effective mass parameter, which lead to the symmetry restoration phenomenon (see, for example, Refs.~\cite{Cline:1995dg, Carena:1996wj,deCarlos:1997ru,Carena:1997ki,Laine:1998qk,Delaunay:2007wb,Carena:2008vj, Cohen:2011ap,Laine:2012jy,Curtin:2012aa,Carena:2012np,Carena:2004ha,Huang:2012wn,
Fok:2008yg,Davoudiasl:2012tu,Gorbahn:2015gxa}). Since these new physics ingredients affect mainly the Higgs potential, it is expected that they will alter the Higgs couplings. One collider observable to test the possibility of the EW phase transition is the triple Higgs coupling~\cite{Huang:2016cjm, Cepeda:2019klc}.

Several simple extensions of the SM are capable of generating the required extra
terms in the potential for a strong first-order EWPT. The simplest extension is to add extra scalar states to the SM. Examples include scalar singlet extensions~\cite{Noble:2007kk,Profumo:2007wc,Barger:2011vm,Wainwright:2012zn,Patel:2012pi,Profumo:2014opa,Katz:2014bha,Kozaczuk:2015owa,Chen:2017qcz,Vaskonen:2016yiu,Carena:2019une}, Two Higgs Doublet Models (2HDMs)~\cite{Bochkarev:1990fx,McLerran:1990zh,Turok:1990zg,Cohen:1991iu,Nelson:1991ab,Basler:2016obg}, the (Next to) Minimal Supersymmetric Standard Model~\footnote{In the MSSM, the 1st order EWPT is excluded by the measured Higgs mass, $m_h = 125$~GeV, and by light stop searches~\cite{Carena:2012np}.} (MSSM/NMSSM)~\cite{Menon:2004wv,Kozaczuk:2014kva}, or Composite Higgs Models~\cite{Creminelli:2001th,Randall:2006py,Nardini:2007me,Grinstein:2008qi,Espinosa:2011eu,Bruggisser:2018mus,Xie:2020bkl}. Recently, it is shown that when the critical temperature is low compared to the new fermion masses, fermions contribute to the Higgs effective potential in the same way as scalars, and therefore can lead to non-trivial effects, including a strong first-order EWPT in the thermal history of the early universe~\cite{Egana-Ugrinovic:2017jib,Angelescu:2018dkk}.

In addition to collider searches, gravitational wave experiments are a promising new probe of cosmological phase transitions~\cite{Caprini:2019egz}, as they can be sensitive to the stochastic gravitational wave signal that gets formed when bubbles of broken electroweak vacuum collide. If the phase transition takes place around the electroweak scale (as is the case in electroweak baryogenesis), space-based interferometers are most sensitive to the signal~\cite{2017arXiv170200786A}. In the case of strong supercooling, observation by ground-based gravitational wave telescopes is even possible~\cite{Baldes:2018emh}. Recent computations of the wall velocity suggest that the parameter space of successful electroweak baryogenesis \emph{and} efficient gravitational wave production is limited~\cite{Cline:2021iff,Lewicki:2021pgr,Dorsch:2021nje}, but alternative models with particle production by relativistic-moving bubble walls could be probed~\cite{Baldes:2021vyz, Azatov:2021irb}.

\section*{Acknowledgements}
The work of R.C. was supported by DoE grant DE-SC0011842 at the University of Minnesota. Y.C. is supported in part by the US Department
of Energy under award number DE-SC0008541. The research of F.E. and G.E. is supported by the Cluster of Excellence {\em Precision Physics, Fundamental Interactions and Structure of Matter\/} (PRISMA${}^+$ -- EXC~2118/1) within the German Excellence Strategy (project ID 39083149). F.E. is also supported by by grant 05H18UMCA1 of the German Federal Ministry for Education and Research (BMBF).
R.M. is supported in part by the DoE grant DE-SC0007859.
The work of GW is supported by World Premier International Research Center Initiative (WPI), MEXT, Japan.  The work of H.D. is supported by the United States Department of Energy under Grant Contract DE-SC0012704.
The work of PH is supported by the National Science Foundation under grant number PHY-1820891, and PHY-2112680, and the University of Nebraska Foundation. A.G. is supported by the U.S. Department of Energy under grant No. DE–SC0007914.
MS is supported by TRIUMF who receives federal funding via a contribution agreement with the National Research Council of Canada. The work of J.H., C.H. and K.F. was supported by the DFG Emmy Noether Grant No. HA 8555/1-1. K.F.
additionally acknowledges support from the DFG Collaborative Research Centre “Neutrinos and DarkMatter in Astro- and Particle Physics” (SFB 1258).
The work of BS and JvdV is supported by the Deutsche Forschungsgemeinschaft
under Germany’s Excellence Strategy - EXC 2121 Quantum Universe - 390833306.

\bibliographystyle{JHEP}
\bibliography{ref}{}

\end{document}